\documentclass[twocolumn]{aastex631}

\usepackage{graphicx}
\usepackage{amsmath}

\usepackage[caption=false]{subfig}

\usepackage{xcolor}
\usepackage{comment}

\begin{document}

\title{An FPCA-Enhanced Ensemble Learning Framework for Photometric Identification of Type Ia Supernovae}

\author{Moonzarin Reza}
\affiliation{Department of Physics and Astronomy, Texas A\&M University} 
\affiliation{George P. and Cynthia Woods Mitchell Institute for Fundamental Physics and Astronomy, Texas A\&M University}

\author{Lifan Wang}

\affiliation{Department of Physics and Astronomy, Texas A\&M University} 
\affiliation{George P. and Cynthia Woods Mitchell Institute for Fundamental Physics and Astronomy, Texas A\&M University}

\author{Lei Hu}
\affiliation{McWilliams Center for Cosmology and Astrophysics,
    Department of Physics,
    Carnegie Mellon University,
    5000 Forbes Avenue, Pittsburgh, PA 15213}

\begin{abstract}

Type Ia supernovae (SNe~Ia) are essential tools for addressing key cosmic questions, including the Hubble tension and the nature of dark energy. Modern surveys are predominantly photometry-based, making the construction of a clean photometric SNe~Ia sample crucial. In this study, we investigate whether functional principal component analysis (FPCA) scores derived from photometric light curves, combined with ensemble learning, can reliably distinguish SNe~Ia from other transients using the PLAsTiCC dataset. FPCA provides a data-driven, flexible characterization of light curves without relying on rigid theoretical model assumptions. Light curves are fitted by minimizing residuals with penalty terms from clean samples, making the method robust to outliers or poorly sampled bands. The first two FPCA scores and peak magnitudes across the five LSST bands are used as classification features. We implement two complementary binary classifiers: an ensemble boosting model (CatBoost) and a statistical probabilistic method based on Euclidean distances. CatBoost slightly outperforms the statistical method, achieving 98.5\% accuracy and 97.8\% precision. Performance remains robust ($>90\%$) under typical photometric redshift uncertainties ($\sigma = 0.1$). On the spectroscopic DES Y5 sample, both methods reach approximately 90\% accuracy and 95\% precision, demonstrating strong out-of-domain generalization compared to state-of-the-art methods with limited cross-survey applicability. Applied to DECam DDF and DESIRT transients, the predictions strongly agree, and their intersection provides a high-confidence SNe~Ia sample for cosmological analyses. Overall, this FPCA-based framework offers a powerful, flexible tool for classifying transients in upcoming large-scale surveys such as LSST and Roman.

\end{abstract}

\keywords{Type Ia Supernovae --- Machine Learning  --- Light Curves --- Principal Component Analysis }

\section{Introduction} \label{sec:intro}

The discovery of the accelerated expansion of the Universe in the late 1990s by the High-Z Supernova Search Team \citep{riess1998observational} and the Supernova Cosmology Project \citep{perlmutter1999measurements}, based on samples of both low- and high-redshift Type Ia supernovae (SNe Ia), established a positive cosmological constant $(\Omega_\Lambda)$ and a negative deceleration parameter ($q_0$). This result revolutionized modern cosmology, revealing that approximately 70\% of the mass–energy budget of a flat Universe today is associated with dark energy. Since then, SNe Ia have become a cornerstone of cosmological studies: with careful calibration of their peak luminosity, a pure sample of SNe Ia provides robust constraints on cosmological parameters.

Large-scale optical surveys such as the Dark Energy Survey \citep{dark2016dark} and Pan-STARRS \citep{chambers2016pan} have already discovered thousands of transients, while upcoming surveys—including the Legacy Survey of Space and Time (LSST; \citealt{sweeney2006overview}) and the Roman Wide Field Imaging Survey \citep{roman2019nancy}—are expected to detect on the order of $10^5$ supernovae. However, the logistical demands of spectroscopic follow-up cannot scale accordingly, as such observations require high sensitivity and significant resources. The transient nature and faintness of distant SNe further exacerbate this challenge. Consequently, there is growing emphasis on developing reliable methods to classify and analyze supernovae using photometric data alone.

In parallel, machine learning has become an indispensable tool across astronomy, with successful applications spanning exoplanets \citep{jin2022identifying,zhao2021machine}, galaxies \citep{reza2021galaxy,reza2020photometric,krakowski2016machine}, stellar populations \citep{zammit2024machine,kuntzer2016stellar}, cosmological simulations \citep{reza2024constraining,reza2024predicting}, and supernovae \citep{chen2022using,chen2024artificial,hu2022spectroscopic}. 

For SNe Ia, the long-standing approach of template fitting \citep{guy2005salt,goldhaber2001timescale} remains widely used in cosmological analyses, but machine learning techniques promise improved classification performance when paired with appropriate models and feature extraction strategies.
Several machine learning studies have already demonstrated strong performance on supernova classification tasks. For example, \citet{santos2020cosmological} benchmarked a variety of feature-extraction and classification methods using the Supernova Photometric Classification Challenge dataset and found that a tree-based pipeline combined with SALT2 features \citep{guy2007salt2} outperformed alternatives. Recurrent neural networks trained on SNANA-simulated light curves achieved accuracies exceeding 96\% for binary classification \citep{moller2020supernnova}. Ensemble-learning approaches combining simulated and observed data from the Supernova Legacy Survey (SNLS) reached an ``Area Under the receiver operating characteristics Curve (AUC)", a single metric which gives a comprehensive view of model performance across different probability cutoffs, of 0.98 \citep{moller2016photometric}. Gaussian process regression has also been widely employed for modeling and interpolating light curves. For example, \citet{qu2021scone} used Gaussian regression to generate flux maps in wavelength–time space, which were subsequently classified by a convolutional neural network, achieving accuracies of 99.7\% (in-distribution) and 80.6\% (out-of-distribution). This framework was later adapted for early supernova classification \citep{qu2022photometric}, achieving 75\% accuracy at the trigger date and 89\% accuracy at 50 days post-trigger. Similarly, \citet{boone2019avocado} used Gaussian process regression to augment training data, extracting features for a LightGBM classifier that achieved an AUC of 0.957. Other approaches include recurrent networks for binary classification \citep{charnock2017deep}, parametric feature extraction combined with gradient-boosted trees for multi-class classification \citep{de2024superphot}, and CNN-based methods trained on heat maps \citep{burhanudin2023pan}.

Generative models have also been explored. For example, \citet{boone2021parsnip} proposed ParSNIP, in which light curves are encoded through one-dimensional convolutional layers into latent variables, concatenated with meta-parameters (reference time, color, and amplitude) and passed to a classifier. However, the complexity of this architecture introduces risks of overfitting and strong dependence on survey-specific characteristics. Although data augmentation strategies were applied to mitigate these effects, the model has not been validated on independent datasets, leaving its generalization power uncertain.

In this work, we propose a machine learning–based framework for classifying astronomical transients in upcoming large-scale surveys, such as those from the Vera C. Rubin Observatory’s LSST \citep{tyson2002large}. We employ functional principal component analysis (FPCA) to represent multi-band light curves with arbitrary sampling rates using two principal components, providing a compact and survey-independent feature representation. The FPCA scores derived for the $grizy$ filters serve as inputs to an ensemble-based gradient-boosting model, which performs binary classification of transients in the PLAsTiCC dataset. We further test the model on observed DECam light curves to evaluate its domain adaptability.

This paper is organized as follows. Section~\ref{sec:dataset} describes the dataset used in this study. Section~\ref{sec:methods} outlines the FPCA model and light-curve fitting procedure, and Section~\ref{sec:machinelearning} presents the machine learning classifier. Results based on PLAsTiCC simulations are reported in Section~\ref{sec:results}, followed by applications to DECam data in Section~\ref{applications}. Conclusions and discussions are given in Section~\ref{sec:conclusions}.

\section{Dataset} \label{sec:dataset}

The Legacy Survey of Space and Time (LSST) will be carried out with the 8.4-m telescope at the Vera C. Rubin Observatory in Chile, designed with a wide field of view to address key questions about the Universe, including the nature of dark matter and dark energy, the formation and structure of the Milky Way, the inventory of the Solar System, and the study of astrophysical transients \citep{gressler2014lsst}.

To prepare for the challenge of classifying LSST transients, the Photometric LSST Astronomical Time Series Classification Challenge (PLAsTiCC) was developed. The dataset consists of simulated light curves, i.e. fluxes measured as a function of time in six bandpass filters spanning the ultraviolet, optical, and infrared \citep{allam2018photometric}. The PLAsTiCC training set contains about 8,000 transients from 14 categories, each with spectroscopic redshifts, reflecting what LSST is expected to observe in its first three years (average spectroscopic redshift $\sim$0.25). The PLAsTiCC test set, by contrast, includes 3.5 million more distant and fainter objects (average photometric redshift $\sim$0.51), of which only a small fraction have spectroscopic observations. In addition to the categories represented in the training set, the test set includes five additional hypothesized classes \citep{kessler2019models}. Approximately 99\% of light curves are drawn from the Wide-Fast-Deep (WFD) survey, which covers $\sim$20,000 deg$^2$ and observes each field twice per week in at least one filter, while the remaining 1\% are from the Deep-Drilling Fields (DDF), covering only $\sim$50 deg$^2$ with higher cadence to reach fainter objects.

In this work, we primarily use the PLAsTiCC test set for both training and evaluation, unless stated otherwise. Our choice is motivated by the novelty of our method—our main goal is to test whether FPCA-based scores can robustly separate Type Ia supernovae from other transients. For comparison, we also trained the model on the PLAsTiCC training set and evaluated on the PLAsTiCC test set, following the original challenge setup. We refer to this latter case as the out-of-distribution test (For a description of the dataset, see  \citet{hlovzek2023results}).

From the dataset, we extract all transients detected within 10 days of maximum light with at least five data points in at least one filter. The resulting dataset contains 11,158 light curves, summarized in Table \ref{tab:data_types}. Of these, 6,048 (54.2\%) are Type Ia supernovae, grouped as “Type Ia,” and 5,110 (45.8\%) belong to other categories, grouped as “Non–Type Ia.” Unlike previous studies (e.g., \citealt{qu2021scone}), which restricted analyses to well-sampled light curves from the DDF survey or to specific transient types, we apply no such cuts. Our sample spans the full redshift range ($0 < z < 3$) and includes objects from both WFD and DDF surveys, covering the entire set of transient types available.

\begin{table}
\centering
\caption{Types of objects in the PLAsTiCC dataset used.}
\label{tab:data_types}
\begin{tabular}{ccc}
\hline
Name & Description & Number\\
\hline
SNIa & Type Ia Supernova & 6048 \\
SNIa-91bg & Peculiar Type Ia Supernova: 91bg & 122 \\
SNIax & Peculiar Type Ia Supernova: SNIax & 275 \\
SNII & Core collapse, Type II supernova & 3921 \\
SNIbc & Core collapse, Type Ibc supernova & 675\\
SLSN-I & Super-luminous supernova (magnetar) & 31\\
TDE & Tidal disruption event &32\\
ILOT & Intermediate luminous optical transient &14 \\
CaRT & Calcium-rich transient &33\\
PISN & Pair-instability supernova&1 \\
$\mu$Lens-Single & $\mu$-lens from single lens &4\\
$\mu$Lens-Binary & $\mu$-lens from binary lens &2\\
\hline
\end{tabular}
\end{table}

\section{Methods} \label{sec:methods}
In this section, we describe the functional principal component analysis (FPCA) model adopted to fit supernova light curves. We also outline the fitting procedure, including the use of the MPFIT module for optimizing model parameters, and provide the detailed steps of our light-curve fitting pipeline.
\subsection{FPCA Model}
Functional principal component analysis (FPCA) is a purely data-driven model for Type Ia supernova (SN Ia) light curves, designed to project sparsely sampled observations onto a set of smooth, complete light curves \citep{he2018characterization}. The basis functions, or principal components, are mathematical constructs chosen to maximize the variance captured in the data. Unlike physics-based models such as SALT2 \citep{taylor2023salt2} or SNooPy \citep{burns2010carnegie}, FPCA relies minimally on fixed assumptions, enabling it to capture more subtle and diverse features of the data.
The principal components used here are derived from a combined sample of well-observed SNe Ia from the Lick Observatory Supernova Search \citep{leaman2011nearby}, the Carnegie Supernova Project \citep{hamuy2006carnegie}, and the Harvard-Smithsonian Center for Astrophysics sample \citep{hicken2012cfa4}. The training sample was restricted to supernovae with at least one observation within five days both before and after maximum light in all filters.
In FPCA, each light curve is expressed as a weighted sum of principal components that characterize its shape. For example, the zeroth-order component represents the mean light-curve shape, the first component encodes light-curve width, and the second component accounts for variations in post-peak decline rate \citep{he2018characterization}.
The FPCA model with two principal components for the light curve in the $i$-th band at observation time $t_{ij}$ is given in Equation \ref{eqn:1}:

  \begin{equation}
\begin{aligned}
f_{model}(t_{ij}, z; \Theta_i) &= \hat{m}_i + \phi_{0}(\frac{t_{ij} - \hat{t}_i}{1 + z})  
\\
&+ \alpha_i \phi_{1}(\frac{t_{ij} - \hat{t}_i}{1 + z}) + \beta_i \phi_{2}(\frac{t_{ij} - \hat{t}_i}{1 + z}), 
\\
\Theta_i &= \{ \hat{m}_i, \hat{t}_i, \alpha_i, \beta_i \},
\end{aligned}
\label{eqn:1}
\end{equation}
  
  Here, $\hat{m}_i$ denotes the peak magnitude in the $i$-th band, $\hat{t}_i$ is the corresponding peak time, and $(\alpha_i, \beta_i)$ are the first two FPCA coefficients for that band. The quantity $t_{ij}$ represents the time of the $j$-th observation in band $i$, and $z$ is the known photometric redshift. The functions $\phi_0$, $\phi_1$, and $\phi_2$ correspond to the zero-, first-, and second-order principal components of the FPCA model, respectively. The model is fully specified by the parameter vector $\Theta_i$ for each band. We employ a filter-vague FPCA model, in which the principal components are identical across all filters. This simplifies the analysis, as the same components can be applied to any broadband filter system in both the rest frame and the observer’s frame.

 \subsection{MPFIT Module}
We fit the light curves using the MPFIT module, which solves non-linear least-squares fitting problems \citep{markwardt2009non} using the Levenberg–Marquardt technique. The model parameters are determined by fitting a user-defined function to the observed data. In this work, we estimate the best-fit parameters through chi-square minimization. The module also provides flexibility by allowing both fixed values and bounded constraints on the parameters. It returns the optimized parameter values along with their covariance matrix.
\subsection{Light-Curve Fitting}
\label{sec:light_curve_fitting}
We fit the FPCA model, represented by $\Theta_i$, to multi-band photometric data of both SN Ia and non–SN Ia transients using MPFIT. Although the FPCA model was originally developed for Type Ia supernova light curves, we extend its application to the full range of transients in our dataset. By using a two-component FPCA model, we retain enough flexibility to capture diverse temporal behaviors  across light curves, while avoiding the rigidity that would result from adding more components. The fitting steps are outlined below.

\textbf{Step 0}: Well-sampled light-curve bands (those with at least five observations) are used to determine the four components of the parameter vector $\Theta_i$ by minimizing chi-square, as shown in Equation~\ref{eq:without_prior}.

\begin{equation}
\tilde{\chi}^2_i = \sum_j \left( \frac{m_{ij} - f_{model}(t_{ij}, z; \Theta_i)}{\sigma_{ij}} \right)^2,
\label{eq:without_prior}
\end{equation}

where $m_{ij}$ and $\sigma_{ij}$ are the measured magnitude and its $1\sigma$ uncertainty for the $j$-th observation in band $i$. This step is performed to derive penalty terms for:
1)  The FPCA coefficients, to reduce the influence of outliers, and
2) The differences in peak times across multiple bands, to enable modeling of light curves with poor time coverage when at least one band is well sampled.

Because this step is used to establish the penalty terms, we restrict the sample to well-sampled light curves for which the parameter uncertainties are reasonably small and the reduced chi-square ($\chi^2/\nu$) is close to unity.
The penalty on the FPCA coefficients is modeled as a linear function of redshift, consistent with the approximately linear dependence observed (see Appendix). Specifically, $b_{1i}$ and $b_{2i}$ represent the redshift-dependent and constant terms, respectively, for the first penalty score in band $i$, while $c_{1i}$ and $c_{2i}$ are the corresponding terms for the second score. These coefficients are determined via least-squares minimization.

The penalty on the FPCA coefficients is modeled as a linear function of redshift, consistent with the observed dependence (see Appendix). Specifically, $b_{1i}$ and $b_{2i}$ represent the redshift-dependent and constant terms, respectively, for the first penalty score in band $i$, while $c_{1i}$ and $c_{2i}$ are the corresponding terms for the second score. These coefficients are determined through least-squares minimization.

\begin{equation}
\begin{aligned}
S^2_i(\alpha_i, \beta_i, z) &=  S'^{2}_{i} (\alpha_i, z) + S''^{2}_{i}(\beta_i, z), \\
S'^{2}_{i}(\alpha_i, z) &= 
\frac{\big[ \alpha_i - (b_{1i} z + b_{2i}) \big]^2}
{\sigma_{b_{1i}}^2 z^2 + \sigma_{b_{2i}}^2}, \\
{S''_i}^2(\beta_i, z) &=
\frac{\big[ \beta_i - (c_{1i} z + c_{2i}) \big]^2}
{\sigma_{c_{1i}}^2 z^2 + \sigma_{c_{2i}}^2}.
\end{aligned}
\label{eqn:prior_scores}
\end{equation}

The interband peak time offsets (between bands $i$ and $k$) are modeled using a redshift-dependent relation, as given in Equation~\ref{eq:prior_times}. Here, $d_{1ik}$ represents the redshift-dependent term, and $d_{2ik}$ is the constant term, which relates the peak time in the well-sampled band $k$ to that in the poorly sampled band $i$.

\begin{equation}
T^2_i(\hat{t}_i, z) =
\frac{\big[ \hat{t}_i - (d_{1ik} z + d_{2ik} + \hat{t}_k) \big]^2}
{\sigma_{d_{1ik}}^2 z^2 + \sigma_{d_{2ik}}^2 + \sigma_{\hat{t}_k}^2}.
\label{eq:prior_times}
\end{equation}

 \textbf{Step 1}: This step is performed only for well-sampled bands (i.e., those with at least five observations) and is used to determine the four components of the parameter vector $\Theta_i$. The residuals in this step account for both the observed data (Equation~\ref{eq:without_prior}) and the penalty terms for the FPCA coefficients (Equation~\ref{eqn:prior_scores}). Incorporating these penalties ensures that the fitted light-curve shapes are physically reasonable, which in turn improves the estimates of the peak times. The resulting equation is:
 
\begin{equation}
\chi^2_{i(\mathrm{well\text{-}sampled)}} 
=  \tilde{\chi}^2_i + S^2_i(\alpha_i,\beta_i,z)
\end{equation}

 \textbf{Step 2:} This step targets poorly sampled bands ($i$), which may contain as few as two observations, while requiring at least one well-measured band ($k$) as reference (see Equation~\ref{eq:prior_times}). The parameter vector $\Theta_i$ for the poorly sampled band is determined by minimizing residuals with respect to the observed data, the FPCA coefficient penalties, and a penalty enforcing consistency of the peak-time offset relative to the well-sampled band. This ensures that even sparsely sampled light curves inherit physically reasonable shapes and temporal alignment, while fully leveraging information from well-measured bands. The resulting equation is:
 
 \begin{equation}
\chi^2_{i(\mathrm{poorly\text{-}sampled)}} = \tilde{\chi}^2_i + S^2_i(\alpha_i, \beta_i, z) + T^2_i(\hat{t}_i, z)
\end{equation}
 
It is worth noting that even after this step, some bands may still have missing parameter values, as many PLAsTiCC light curves lack observations in certain filters. Consequently, handling missing data remains a key challenge in this work, requiring an appropriate modeling approach.

\begin{figure}
    \centering
    
        \includegraphics[width=\linewidth]{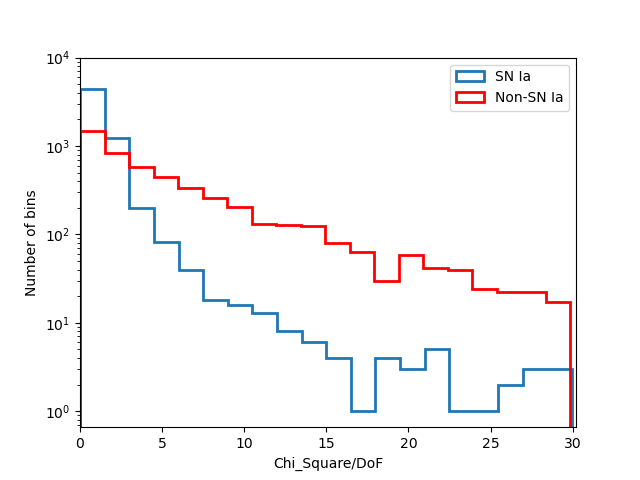}
    
    \caption{Distribution of reduced chi-square ($\chi^2/\nu$) for Type Ia (blue) and Non-Type Ia (red) supernovae. Since the FPCA model used to fit all transients is specifically designed for SNe Ia, the goodness-of-fit, measured by $\chi^2/\nu$, is on average lower (indicating a better fit) for Type Ia than for Non-Type Ia transients.}
    \label{fig:chi_dis}
\end{figure}

We conclude this section by examining the goodness-of-fit through the distribution of the reduced chi-square, $\tilde{\chi}^2_\nu$ (defined as $\tilde{\chi}^2_\nu = \chi^2 / \nu$,  where $\nu$ is the number of degrees of freedom), shown in Figure~\ref{fig:chi_dis}. The blue histogram corresponds to fitted Type Ia supernovae (SNe Ia), while the red histogram represents all other transient types. As expected, the FPCA model is specifically designed for SNe Ia, resulting in systematically lower $\tilde{\chi}^2_\nu$ values for Type Ia objects (blue) compared to Non-Ia transients (red).

\section{Machine Learning Model}  
\label{sec:machinelearning}

In this section, we describe the machine learning model, Categorical Boosting (CatBoost), used in this study, followed by the performance metrics employed to evaluate it. We choose CatBoost because it can naturally handle missing values, which are common in light curves lacking observations in certain bands.

\subsection{Categorical Boosting}

Categorical Boosting \citep{ibrahim2020comparison} is an ensemble learning method whose base learners are decision trees. Each tree splits the data at nodes based on features to minimize impurity, with leaf nodes assigning class labels. Trees are built sequentially, with each tree correcting the errors of previous trees by assigning higher weights to misclassified samples. Predictions from multiple trees are combined using weighted probabilities for classification, minimizing cross-entropy, which has been suggested as an optimal metric for challenges like PLAsTiCC, which needs to baalnce diverse science goals \citep{malz2019photometric}. The model also provides a confidence probability based on the fraction of trees agreeing on the predicted class. Unlike bagging \citep{breiman1996bagging, yu2021bagging}, where trees are constructed independently and predictions rely on majority voting, CatBoost’s sequential approach improves performance on challenging datasets. CatBoost classifiers have been successfully applied in astronomy \citep{khramtsov2019kids}, medicine \citep{dhananjay2021analysis}, cybersecurity \citep{8867696}, and digital marketing \citep{KOEHN2020113342}.

\subsection{Input Features}

For each transient, we construct input features from the FPCA light-curve fits across all five LSST bands (g, r, i, z, y). We use the first two FPCA coefficients in each band, which capture the dominant temporal variations, along with the peak absolute magnitude derived from the fitted peak apparent magnitude assuming a flat $\Lambda$CDM cosmology. These features provide a compact representation of light-curve shape and intrinsic brightness, forming the basis for classification. We also tested including FPCA fit uncertainties, but found no significant improvement in classification performance; hence, these terms are not included in the final model.

\subsection{Performance Metrics}  
\label{sub:per_met}

We evaluate classifier performance using several standard metrics:

\textbf{Confusion Matrix:} A confusion matrix (Table~\ref{tab:conf_mat}) compares predicted labels against true labels:
\begin{itemize}
    \item \textbf{True Positive (TP):} Correctly classified Type Ia SNe.  
    \item \textbf{False Positive (FP):} Non-Type Ia misclassified as Type Ia.  
    \item \textbf{False Negative (FN):} Type Ia misclassified as Non-Type Ia.  
    \item \textbf{True Negative (TN):} Non-Type Ia correctly classified as Non-Type Ia.  
\end{itemize}

The derived metrics are calculated as:
\begin{align*}
\text{Accuracy} &= \frac{TP + TN}{TP + FP + TN + FN} \\
\text{Precision} &= \frac{TP}{TP + FP} \\
\text{Recall} &= \frac{TP}{TP + FN} \\
\text{F1 Score} &= \frac{2 \cdot \text{Precision} \cdot \text{Recall}}{\text{Precision} + \text{Recall}}
\end{align*}

\textbf{ROC and AUC:} The Receiver Operating Characteristic (ROC) curve plots the True Positive Rate ($TPR = TP/(TP+FN)$) against the False Positive Rate ($FPR = FP/(FP+TN)$) across varying threshold probabilities, providing a comprehensive assessment of classification performance. The Area Under the ROC Curve (AUC) quantifies overall classifier performance; a value of 1 indicates a perfect classifier, 0.5 corresponds to random guessing, and values below 0.5 suggest potential systematic issues. Higher AUC indicates better model performance, especially in imbalanced datasets where overall accuracy may be misleading.

 \begin{table}
\centering
\caption{Confusion matrix showing the number of test samples correctly and incorrectly classified by the model. True Positives (TP) and True Negatives (TN) indicate correct classifications, while False Positives (FP) and False Negatives (FN) indicate misclassifications.}

\label{tab:conf_mat}
\begin{tabular}{lccc} 
\hline
& \hspace{2.1em} Predicted \hspace{1.1em}&  Label\\
\hline
& Positive & Negative\\
 True  \hspace{1.1em}\vline \hspace {1em} Positive & TP & FP\\
 Label \hspace {0.735em}\vline \hspace {1em} Negative & FN & TN\\
\hline

\end{tabular}
\end{table}

\begin{figure*}
    \centering
    
        \includegraphics[width=\linewidth]{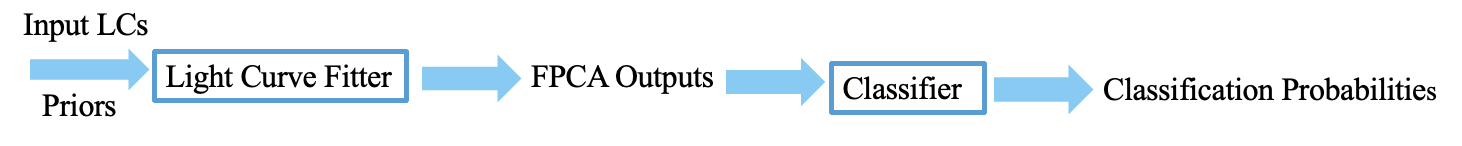}
    
    \caption{This block diagram illustrates the analysis framework employed in this study. Photometric light curves, together with associated penalty terms, are first fitted using the FPCA model. The resulting FPCA features are then used as input to the machine learning classifier, which outputs the predicted probabilities that each sample belongs to different classes, such as Type Ia and Non-Type Ia.}
    \label{fig:block_diag}
\end{figure*}

We conclude this section by presenting a schematic overview of the process described in Sections~2 and~3 (Figure~\ref{fig:block_diag}). The outputs of the FPCA model, obtained from light-curve fitting, are used as input features for the classifier, which then produces the predicted probabilities for each class, such as Type Ia and Non-Type Ia.

\section{Results} \label{sec:results}

In this section, we present the detailed results of our analysis. We begin with the light-curve fits, followed by an investigation of feature importance, and conclude with an evaluation of the classification accuracy. In addition, we assess how the model’s performance depends on the accuracy of the photometric redshift estimates. Since photometric redshifts are inherently subject to uncertainties, we introduce varying levels of Gaussian noise to the reported redshift values in order to test how much uncertainty the model can tolerate before its performance degrades. We further evaluate the classification accuracy for distinguishing other types of transients, thereby exploring the flexibility of the FPCA representation. Finally, we perform additional classification tests using a statistical method as a point of comparison.

Table \ref{tab:per_LCs} summarizes the distribution of light curves with varying numbers of observed bands. Approximately 30\% of the light curves contain sufficient observations in all five bands to reliably constrain the PCA scores. The fractions of light curves with sufficient observations in 4, 3, 2, and 1 bands are 33\%, 23\%, 11\%, and 3\%, respectively.

\begin{table}
\centering
\caption{Percentage distribution of light curves with sufficient observations in different numbers of filters for the full dataset.}
\label{tab:per_LCs}
\begin{tabular}{cc} 
No. of available filters & \% of light curves  \\
\hline

5 &  30.4\\
4 &  33.2 \\
3 &  23.3 \\
4 &  10.6\\
1 &  2.5\\

\hline
\end{tabular}
\end{table}

\subsection{Light Curve Fitting}
\begin{figure*}
\centering
{\includegraphics[width=\textwidth]{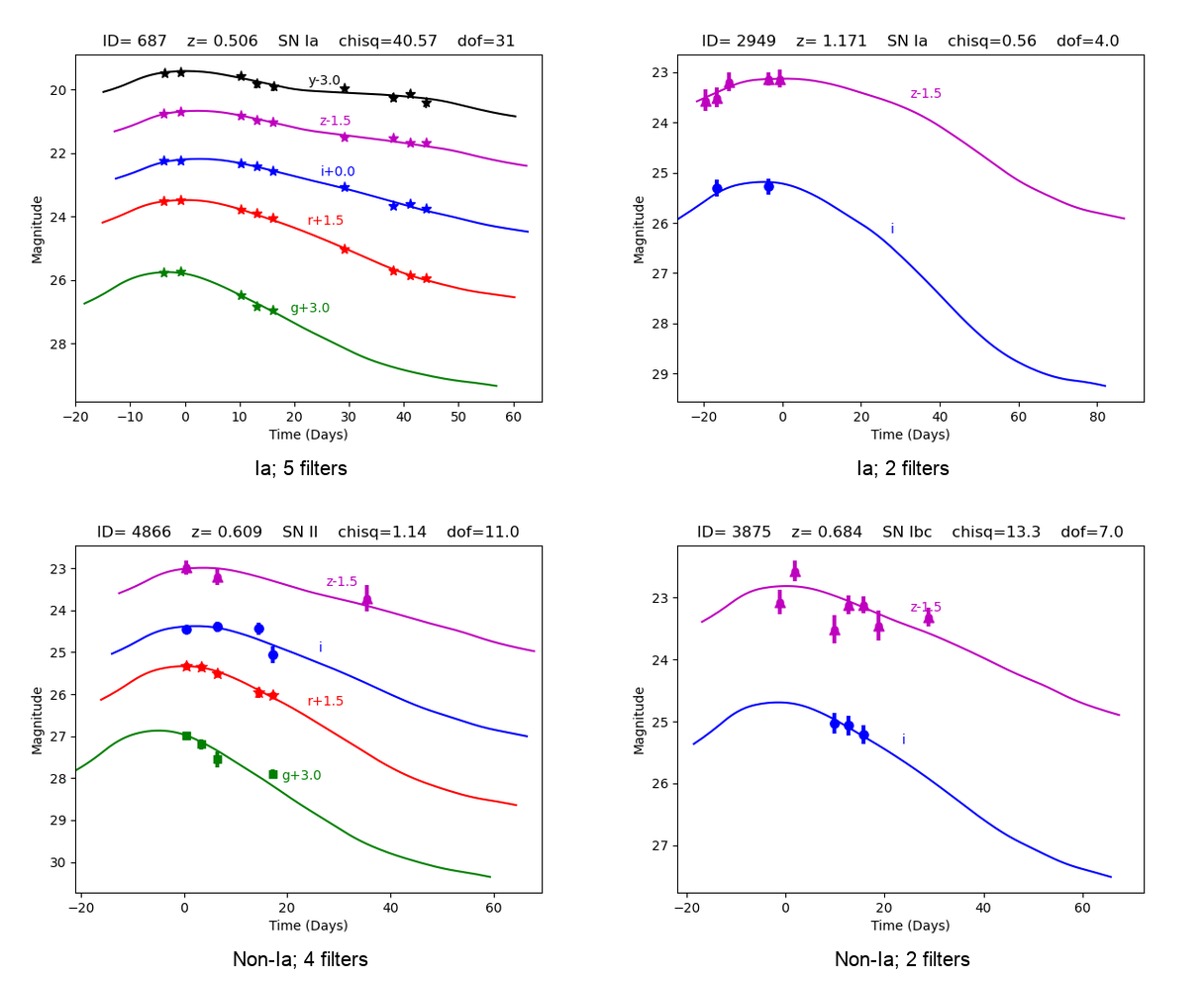}}

\caption{Examples of four light-curve fits with varying numbers of observed filters. The top panel shows two Type Ia supernovae, while the bottom panel presents two core-collapse supernovae.}
\label{fig:example_lc_fits}
\end{figure*}

In this subsection, we describe the light-curve fitting procedure. Four parameters are required to reconstruct the light curves: the time of peak brightness and the peak magnitude, which control the horizontal and vertical shifts, respectively, and the two principal component scores, which determine the  shape  of the light curve. The fitting code is run on an 8-core CPU and requires approximately one hour to process $\sim$\,11,000  light curves across all three steps described in Section 3.2.
Figure \ref{fig:example_lc_fits} shows representative light-curve fits. The upper panel presents two Type Ia supernovae at redshifts 0.506 (left; 5-band coverage) and 1.171 (right; 2-band coverage). Here, t=0 corresponds to the epoch of peak brightness in the band with the maximum number of observations. The fitted curves capture the characteristic features expected for SNe Ia. The lower panel shows examples of Non-Ia transients: a Type II supernova at redshift 0.609 (left; 4-band coverage) and a Type Ibc at redshift 0.684 (right; 2-band coverage).

\subsection{Distribution of FPCA scores}

\begin{figure*}
    \centering
        \includegraphics[width=0.9\linewidth]{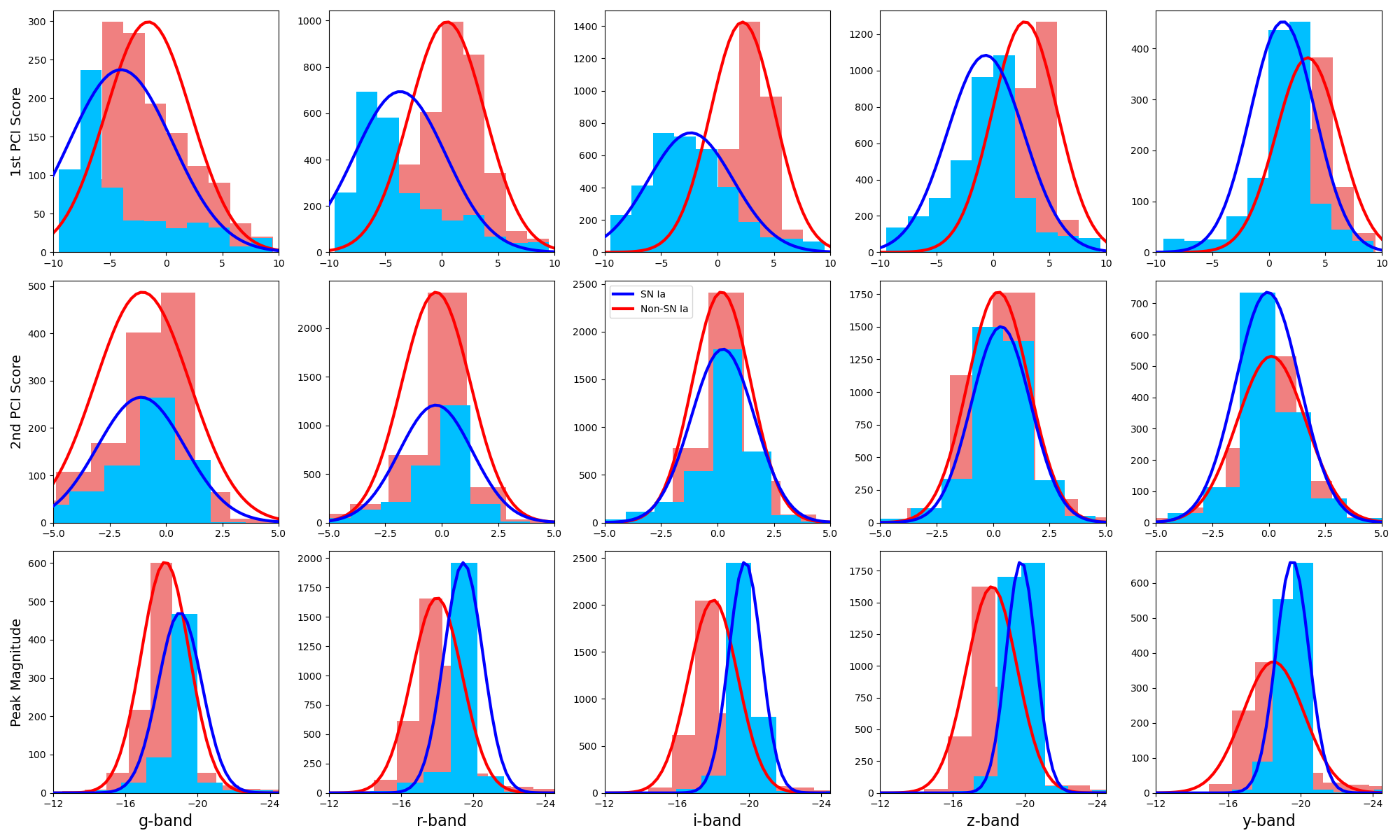}       
    \caption{Histogram distributions with corresponding Gaussian fits for Type Ia supernovae (blue) and non-Type Ia supernovae (red) across fifteen input parameters for the full dataset. The top, middle, and bottom panels show the first principal component scores, second principal component scores, and peak absolute magnitudes, respectively.}
    \label{fig:gau_hist_all}
\end{figure*}

\begin{figure*}
    \centering
        \includegraphics[width=0.9\linewidth]{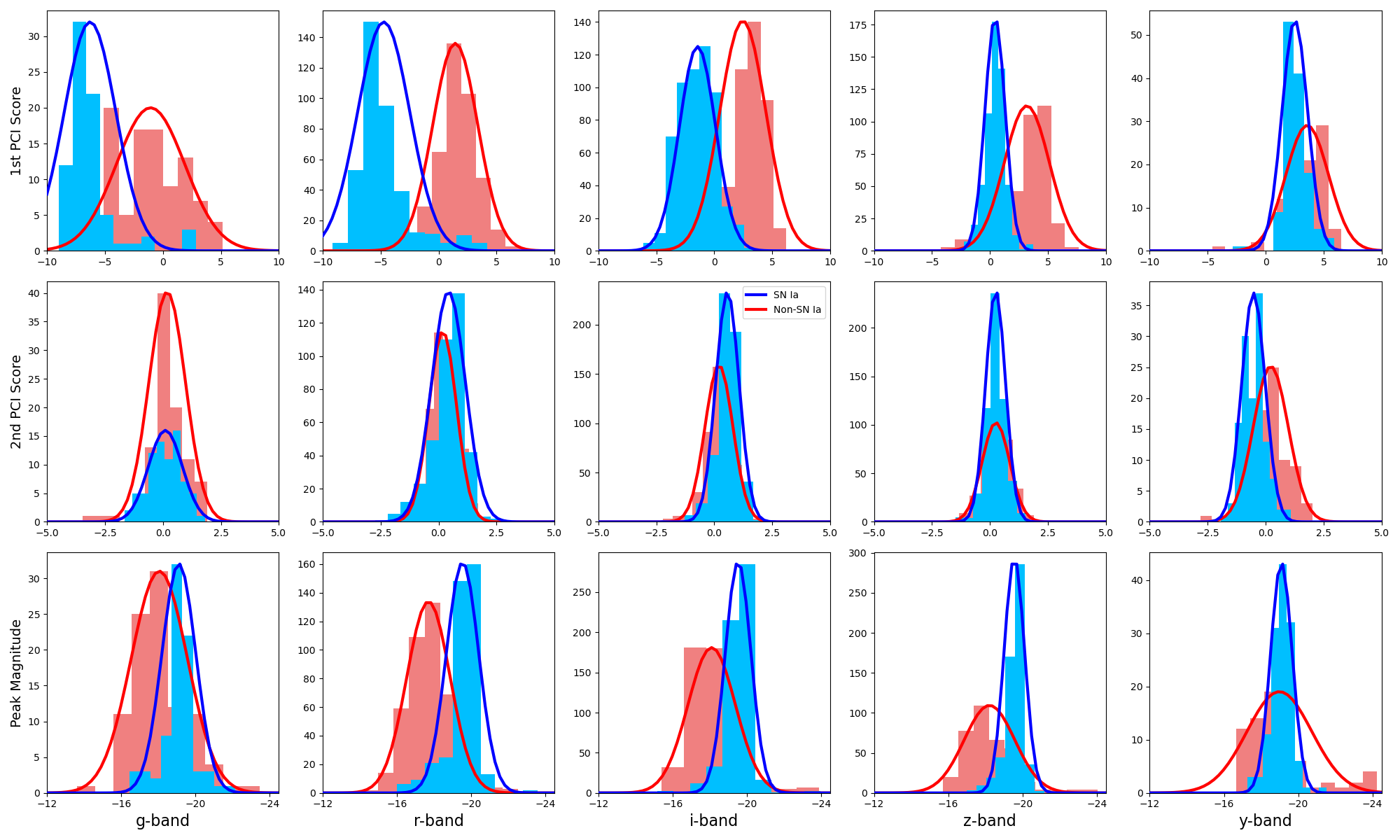}       
          
    \caption{Histogram distributions with corresponding Gaussian fits for Type Ia supernovae (blue) and non-Type Ia supernovae (red) for samples with small parameter uncertainties and well-constrained light-curve fits. The top, middle, and bottom panels show the first principal component scores, second principal component scores, and peak absolute magnitudes, respectively.}
    \label{fig:gau_hist_good}
\end{figure*}

In Figure \ref{fig:gau_hist_all}, we show the histogram distributions of the fitted parameters along with their corresponding Gaussian fits. The blue and red histograms correspond to SNe Ia and all other transients, respectively. The top, middle, and bottom panels display the first principal component score, the second principal component score, and the peak absolute magnitude, with each column representing one photometric band. (The peak apparent magnitudes obtained from light-curve fitting are converted to peak absolute magnitudes using a flat $\Lambda$CDM cosmology.) These distributions reveal a substantial degree of overlap between the two classes, highlighting the need for a non-linear classifier capable of learning the distinguishing patterns directly from the data.
To investigate the effect of data quality on parameter separability, Figure \ref{fig:gau_hist_good} shows the same distributions after discarding samples with large model-fitting errors and high uncertainties in the estimated parameters. Although some contamination remains, it is significantly reduced compared to Figure \ref{fig:gau_hist_all}. The Gaussian peaks of SNe Ia (blue) and Non-Ia transients (red) now exhibit clearer separation, indicating improved discriminative power in this cleaner subset of the data.

\begin{figure*}
    \centering
   
        \includegraphics[width=\textwidth]{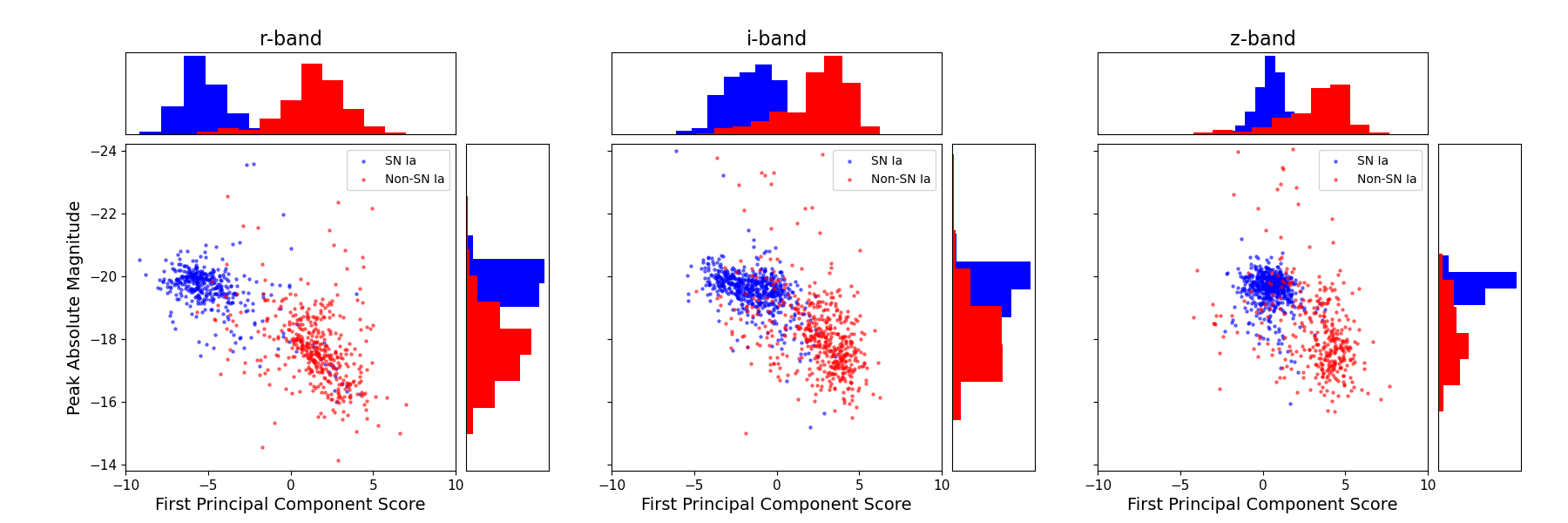}    
    \caption{This figure shows the peak absolute magnitude versus the first principal component score in the r, i, and z bands for samples with small parameter uncertainties and well-constrained light-curve fits. In this feature space, Type Ia and non-Type Ia supernovae exhibit the highest separation. For all bands, the same trend is observed: on average, Type Ia supernovae have higher peak luminosities and lower first principal component scores compared to non-Type Ia supernovae. Additionally, for Type Ia supernovae, the first principal component scores increase progressively from r to i to z bands.}
    \label{fig:dis_param_ori_good}
\end{figure*}

In Figure \ref{fig:dis_param_ori_good}, we show the scatterplot of the peak absolute magnitude (y-axis) vs the 1st PCA score (x-axis), (along with the 1-D histogram) for the r, i and z-bands for only those samples for which the errors in the estimated parameters are reasonable small, with low model-fitting errors.  Type Ia SNe are shown in blue, and other types are shown in red. The reason for choosing these three bands and these two features (not the PC2 score) is that these two classes of transients (Type Ia vs Non-Type Ia) show the highest separation in this parameter space.  For all of these bands, we observe the same trend-- on average the peak luminosity of type  Ia is higher and the PCI score is lower for Type Ia compared to those for Non-Type Ia. Also as we move to longer wavelengths (r-i-z), the first PCA scores show an increasing trend for Type Ia SNe.

\subsection{Feature Importance}
\subsubsection{Impurity based}

In this section, we explore the default feature importance functionality of the CatBoost model in scikit-learn, which is calculated by minimizing cross-entropy \citep{bisong2019introduction}. The basic idea is that features appearing near the root nodes of decision trees are deemed more important than those appearing deeper in the tree. This hierarchy is chosen to minimize impurity at each split for a given feature. For example, if features A and B reduce impurity at the root node by 60\% and 50\%, respectively, feature A is preferentially selected and considered more important. For ensemble methods like CatBoost, feature importance is first calculated for individual trees in this manner and then averaged across all trees to obtain the overall feature importance.

Figure~\ref{fig:feaimp} (left panel) shows the resulting feature importance, with the x-axis representing the importance as a percentage. As expected, absolute magnitudes dominate four out of the top five features, reflecting the well-known precision of Type Ia peak luminosities. The remaining top feature is the first PC score in the \(r\)-band.

\subsubsection{Permutation Importance}

\begin{figure*}
    \centering
  \includegraphics[width=\textwidth]{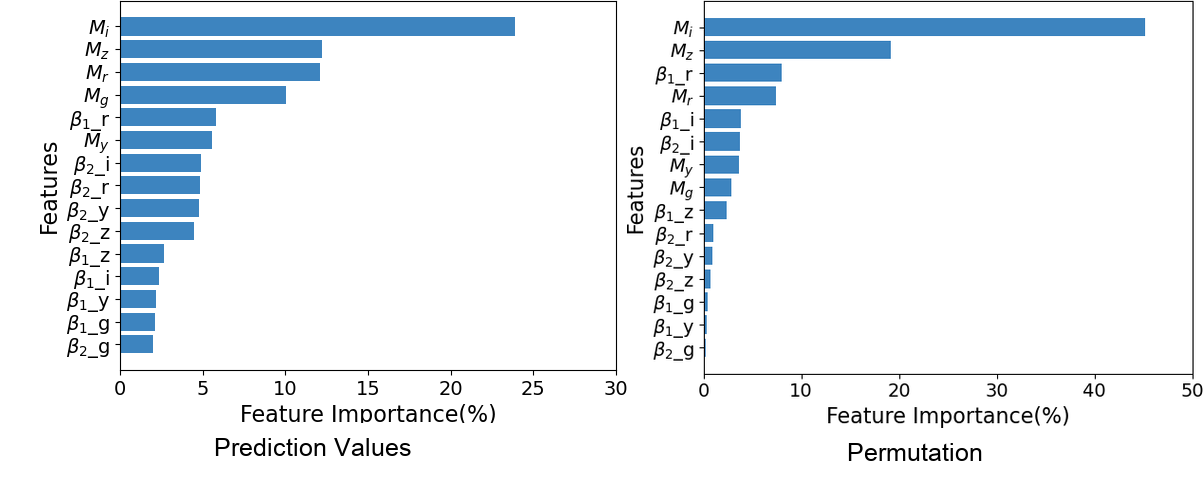}
 
    \caption{Feature importance evaluated using two different methods. The peak absolute magnitudes in the $r$, $i$, and $z$ bands, along with the first PC score in the $r$-band, are the most dominant features according to both models.}

    \label{fig:feaimp}
\end{figure*}

Permutation importance provides a model-independent measure of feature importance that can be applied to any model \citep{altmann2010permutation}. It is computed by shuffling one feature at a time and measuring the resulting increase in the model’s prediction error. Features whose shuffling causes a larger increase in error are deemed more important. In our analysis, we set \(n\_repeats=10\), repeating the shuffling process ten times and averaging the resulting increases in prediction error. While this approach is more computationally intensive than CatBoost’s built-in feature importance, it provides a model-agnostic check. Figure~\ref{fig:feaimp} (right panel) shows the permutation importance, with the x-axis indicating the decrease in accuracy (equivalent to the increase in model error).

Comparing the two panels of Figure~\ref{fig:feaimp}, we find that four of the top five features—peak magnitudes in the \(r\), \(i\), and \(z\) bands, and the first PC score in the \(r\)-band—are consistent across both methods. The fifth feature differs slightly, with the first PC score in the \(i\)-band appearing in permutation importance. Overall, these results indicate that very different methods consistently identify the same features as the most important, and the prominence of peak absolute magnitudes reinforces the model's implicit recognition of Type Ia supernovae as standard candles.

 \label{sec:full_dataset}

\subsection{Model Training and Performance on the Main Dataset}

In this section, we train the model using the PLAsTiCC test set (hereafter referred to as the ``main dataset'' or simply ``dataset'').  The dataset is split into training, validation, and test sets in a 70:10:20 ratio. The input features for the model consist of the first two principal component scores and the peak magnitudes in the available bands.  

Optimal hyperparameters are determined using the validation set, with the prime hyperparameter values summarized in Table~\ref{tab:opt_hyp}.  The classifier is trained five times using different random seeds, and the mean and 1-$\sigma$ uncertainties of the performance metrics across these runs are reported in Table~\ref{tab:met_in_dis}.  

\begin{figure}
  \centering
  
    \includegraphics[width=.9\linewidth]{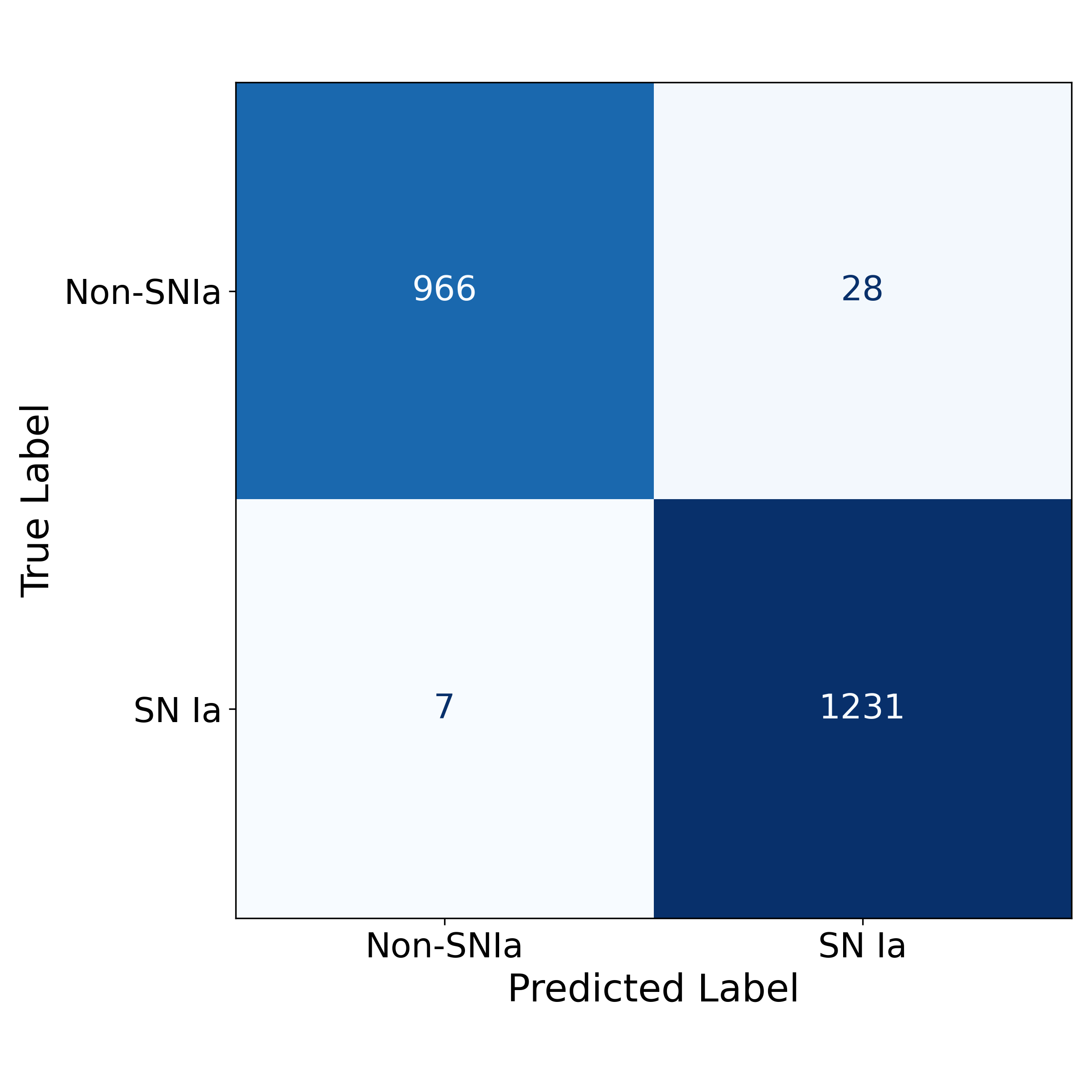}
    \caption{Confusion matrix for the full dataset. Rows correspond to true classes and columns to predicted classes for binary classification. Out of 1,238 Type Ia supernovae in the test set, 1,231 are correctly classified, while out of 994 non-Type Ia supernovae, 966 are correctly classified in this run.}
    \label{fig:cm_no_band}
\end{figure}

The average training, validation, and test accuracies are 99.6\%, 97.0\%, and 98.5\%, respectively. For the test set, the mean precision, recall, and F1-score are 97.8\%, 99.5\%, and 98.6\%, respectively. Figure~\ref{fig:cm_no_band} shows the confusion matrix for one representative run. Out of 1,238 Type Ia supernovae in the test set, 1,231 are correctly classified, and out of 994 non-Type Ia supernovae, 966 are correctly classified for this particular run.

\begin{table}
\centering
\caption{Optimal hyperparameter values for the CatBoost classifier used in this study.}
\label{tab:opt_hyp}
\begin{tabular}{lc} 
Hyperparameter & Value  \\
\hline

iterations &  1000\\
depth &  6 \\
loss\_function &  Logloss \\
learning\_rate &  0.053\\
max\_leaves &  64\\
\hline
\end{tabular}
\end{table}

\begin{table*}
\centering
\caption{Performance on in-distribution data (samples drawn from the same distribution as the training set). Shown are precision, recall, and F1-score for the training, validation, and test sets. For the test set, the mean precision, recall, and F1-score are 97.8\%, 99.5\%, and 98.6\%, respectively.}
\label{tab:met_in_dis}
\begin{tabular}{lcccc} 
Metric & Training(\%) & Validation(\%) &Test(\%)  \\
\hline

Accuracy(\%) &  99.6 $\pm$ 0.1 & 97.0 $\pm$ 0.0 & 98.5 $\pm$ 0.0 \\
Precision(\%) &   99.5 $\pm$ 0.2 & 96.0 $\pm$ 0.1 & 97.8 $\pm$ 0.1 \\ 
Recall(\%) & 99.9 $\pm$ 0.1 & 98.5 $\pm$ 0.1 & 99.5 $\pm$ 0.1  \\
F1-score(\%) &  99.7 $\pm$ 0.1 & 97.2 $\pm$ 0.0 & 98.6 $\pm$ 0.0\\
\hline
\end{tabular}
\end{table*}

\subsection{ROC and AUC}

\begin{figure}
  \centering
  
    \includegraphics[width=.9\linewidth]{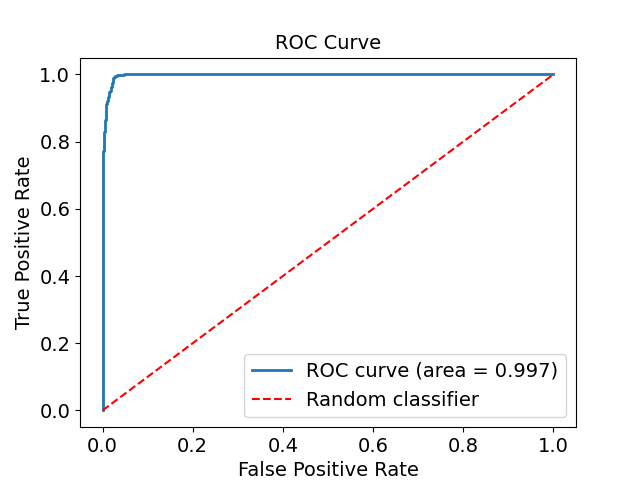}
  \caption{True positive rate versus false positive rate (receiver operating characteristic). The area under the curve (AUC) is 0.997, indicating excellent classification performance.}
  \label{fig:roc}
  \end{figure}

Figure~\ref{fig:roc} shows the receiver operating characteristic (ROC) curve, as defined in Section~\ref{sub:per_met}. The area under the curve (AUC) provides a single measure of the classifier's performance across all threshold probabilities, with values closer to unity indicating better discrimination. The proposed model achieves an AUC of 0.997, demonstrating excellent ability to distinguish Type~Ia supernovae from other transients in the dataset. In general, classifiers with AUC values above 0.9 are considered highly effective.

  \subsection{Threshold probability}

\begin{figure}
    \centering
    \includegraphics[width=.9\linewidth]{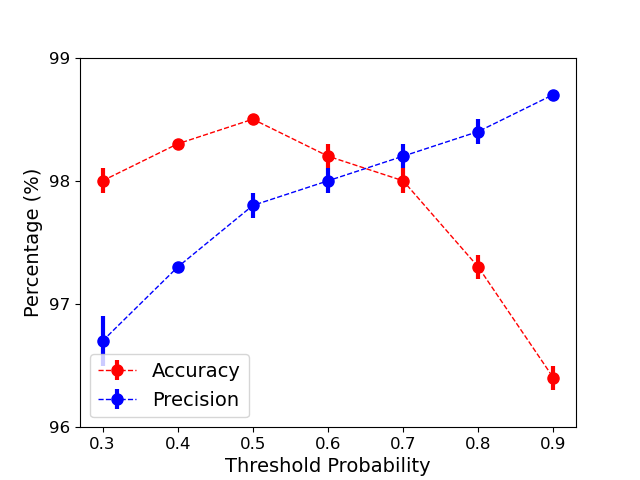}
  \caption{ Accuracy and precision for different threshold probabilities. Increasing the threshold probability monotonically increases precision, whereas accuracy reaches a maximum at a threshold of 0.5 and declines beyond this point, illustrating the trade-off between purity and completeness.}
   \label{fig:thres}
  \end{figure}

  One of the key outputs of ensemble-based learning methods is the confidence probability, defined as the fraction of constituents (here, decision trees) that agree on the model's final prediction. By setting a threshold probability (default 0.5), a sample is classified as Type~Ia if its confidence exceeds the threshold, and as Non-Type~Ia otherwise. This is particularly important for cosmological applications, where obtaining a pure sample of Type~Ia supernovae is critical. Figure~\ref{fig:thres} shows the accuracy and precision for different threshold probabilities. Increasing the threshold monotonically increases precision (blue curve), as stricter criteria yield a purer sample. The overall accuracy (red curve) increases when the threshold rises from 0.3 to 0.5, but decreases beyond 0.5 due to more Type~Ia supernovae being misclassified as Non-Type~Ia, illustrating the trade-off between purity and completeness. Specifically, raising the threshold from 0.5 to 0.9 increases precision from 97.8\% to 98.7\%, while decreasing accuracy from 98.5\% to 96.4\%.

  \subsection{Potential reasons for misclassifications}
 In this section, we explore potential reasons for misclassifications. Table~\ref{tab:mis_z} reports the number of incorrectly classified transients as a function of redshift. For low redshifts ($z \lesssim 0.4$), only 2 SNe~Ia and 1 Non-SNe~Ia are misclassified. One of the SNe~Ia has a light curve spanning only 18 days, with fewer than 10 days past maximum, preventing the FPCA scores from capturing the decline rate accurately. The other SNe~Ia exhibits fluctuations in the $z$- and $y$-band light curves rather than a smooth decline, suggesting that measurement uncertainties are likely responsible.   For the high-redshift transients,  the observed optical band corresponds to the rest-frame ultraviolet (UV), where SALT templates are not well defined \citep{pierel2018extending}.

\begin{table}
\centering
\caption{No. of misclassified transients with redshifts.}
\label{tab:mis_z}
\begin{tabular}{ccccc} 
Redshift range & SNe Ia & Non-SNe Ia  \\
\hline

$0.0<z<0.2$  &   1 & 1 \\
$0.2<z<0.4$ &    1  & 0  \\ 
$0.4<z<0.6$ & 2 & 7  \\

$0.6<z<0.8$ & 2  & 9\\
$0.8<z<1.0$ & 1 &7\\
$z>1.0$ & 0 &8\\

\hline
\end{tabular}
\end{table}

\subsection{Introducing redshift noise}

Large-scale surveys such as LSST and Roman will not be able to obtain accurate spectroscopic redshifts for all detected transients. Measured photometric redshifts are subject to uncertainties from photometric errors, limited wavelength coverage, calibration issues, and other factors. To assess the impact of these uncertainties on our classifier, we perturb the photometric redshifts of the test set as $z' \rightarrow \max(0, z + \mathcal{N}(0, \sigma))$, where $z$ is the quoted photometric redshift, while keeping the training redshifts noise-free. The modified redshift is then used in light-curve fitting. We test $\sigma$ values of 0.1, 0.2, and 0.3, comparing the model's performance to the base classifier (no noise). Each experiment is repeated three times per $\sigma$, with results summarized in Table~\ref{tab:met_red_noise}.  

The overall accuracy of the base classifier is 98.5\%, while it drops to 92.4\%, 82.5\%, and 72.6\% for Gaussian noise levels of $\sigma=0.1, 0.2,$ and $0.3$, respectively, indicating an approximate 10\% decrease in accuracy for every 0.1 increase in noise. Precision remains above 90\% for all noise levels, demonstrating that the contamination of the predicted sample is relatively insensitive to redshift errors. Recall exceeds 90\% for $\sigma=0.1$, but decreases by 18\% and 35\% for larger $\sigma$ values, indicating that higher redshift uncertainties make it increasingly difficult to correctly classify Non-Type~Ia supernovae. The tolerance level thus defines the acceptable noise in the data, which in turn constrains the observational depth, since high-redshift measurements generally carry larger uncertainties. Importantly, at $\sigma=0.1$, which is typical for LSST photometric redshifts, all performance metrics remain above 90\%, indicating that the classifier is robust and reliable at this noise level.

\begin{table*}
\centering
\caption{Performance of the model (accuracy, precision, recall) for different levels of photometric redshift noise.}
\label{tab:met_red_noise}
\begin{tabular}{ccccc} 
 Metric & $\sigma$=0.0 & $\sigma$=0.1 &$\sigma$=0.2& $\sigma$=0.3 \\
\hline
Accuracy(\%) & 98.5 $\pm$ 0.0& 92.4 $\pm$ 0.2 &82.5$\pm$0.3&72.6$\pm$0.1\\
Precision(\%) & 97.8 $\pm$ 0.1 & 95.7$\pm$ 0.1 & 94.6$\pm$ 0.2 & 92.1$\pm$ 0.8\\
Recall(\%) & 99.5 $\pm$ 0.1 & 90.5 $\pm$ 0.2 & 72.6$\pm$ 0.9 & 55.5$\pm$ 0.2 \\
F1-score(\%) & 98.6 $\pm$ 0.0 & 93.0 $\pm$ 0.1 & 82.1$\pm$ 0.4 & 69.3$\pm$ 0.1  \\
\hline
\end{tabular}
\end{table*}

\subsection{Out-of-Distribution}
In this section, we train a CatBoost classifier using the original PLAsTiCC training set (spectroscopic sample) and evaluate it on the original PLAsTiCC test set, which has been used throughout this study. After applying the same constraints used to extract the original dataset, the training set contains 1783 transients. The results are summarized in Table~\ref{tab:met_out_dis}.

 We run the evaluation for five different random seeds and report the mean values of accuracy, precision, recall, and F1-score in Table~\ref{tab:met_out_dis}. The averaged metrics are: accuracy 81.2\%, precision 79.8\%, recall 87.4\%, and F1-score 83.4\%. 

Comparing these out-of-distribution results with in-distribution performance, we observe decreases of 17\%, 18\%, and 15\% in accuracy, precision, and recall, respectively. These discrepancies likely arise from several factors: (i) the training data does not adequately reflect the distributions in the test set \citep{allam2018photometric}, particularly due to differences in the redshift distributions (with mean redshifts of 0.25 for the brighter PLAsTiCC training sample versus 0.45 for the fainter PLAsTiCC test set), (ii) novel objects in the test set (ILOT, CaRT, PISN, $\mu$Lens-Binary; see Table~\ref{tab:data_types}) that are absent from the training set, (iii) no separate penalty terms have been derived for the spectroscopic sample, and (iv) the relatively small training set (1783 samples) limits achievable accuracy.

For comparison, \citet{qu2021scone} report accuracies of 99.7\% and 80.6\% for in-distribution and out-of-distribution sets, respectively. In our study, the drop in accuracy is slightly smaller (17\% vs 19\%), indicating that the FPCA-based approach used here performs comparably to the heatmaps and Gaussian process regression employed in that work. Additionally, unlike \citet{qu2021scone}, we do not apply any cuts based on transient type, time coverage, or signal-to-noise ratio, aiming to develop a model robust enough for real-time classification of LSST light curves. 

We also tested the model on a subset of light curves with sufficient coverage (at least five observed data points in all five filters). Results, reported in the last column of Table~\ref{tab:met_out_dis}, show all performance metrics above 90\%, with mean accuracy and precision of 92.5\% and 94.8\%, respectively. These results demonstrate that satisfactory classification performance can be achieved when light curves have adequate coverage, even if the training and test set statistics differ.

\begin{table*}
\centering
\caption{Performance metrics for the model on out-of-distribution data, i.e., test samples drawn from a distribution with different properties than the training set. This evaluation highlights the model’s ability to generalize beyond the conditions represented in training.}
\label{tab:met_out_dis}
\begin{tabular}{ccccc} 
Metric & In-distribution & Out-of-distribution(all) & Out-of-distribution(good coverage)   \\
\hline
Accuracy(\%) &  98.5 $\pm$ 0.0 & 81.2 $\pm$ 0.9  & 92.5 $\pm$ 0.2 \\
Precision(\%) &   97.8 $\pm$ 0.1 &  79.8$\pm$ 1.3 & 94.8 $\pm$ 0.3 \\ 
Recall(\%) & 99.5 $\pm$ 0.1 &  87.4$\pm $1.4 & 92.7 $\pm$ 0.4 \\
F1-score(\%) &  98.6 $\pm$ 0.0 & 83.4 $\pm$ 0.7 & 93.7 $\pm$ 0.2\\
\hline
\end{tabular}
\end{table*}

\subsection{Multiclass Classification}

 In this section, we investigate the performance of the classifier for multiclass classification across all 10 types of transients in the dataset (see Table~\ref{tab:data_types}). Two classes (PISN and Binary system $\mu$-lensing) are removed due to having fewer than three samples. The resulting dataset remains highly imbalanced, with fewer than five occurrences for Point source $\mu$-lensing and fewer than 100 transients for four additional classes. To address this, we perform class-balanced classification using the Synthetic Minority Oversampling Technique (SMOTE; \citealt{chawla2002smote}).  

SMOTE balances imbalanced datasets by generating synthetic examples of minority classes. Instead of duplicating existing samples, it interpolates between a sample and its nearest neighbors to create new synthetic data points. This procedure helps the classifier learn better decision boundaries and reduces bias toward majority classes. For our analysis, we generate 6048 samples for each class, matching the number of Type Ia SNe, which is the majority class. The resulting augmented dataset is then split into training, validation, and test sets in a 70:10:20 ratio, consistent with the previous analysis.  

We repeat the analysis five times using different model initializations. The full performance metrics are presented in Table~\ref{tab:cat_class}.  The mean test accuracy, precision, and recall for the 10-class problem are 98.3\%, 98.3\%, and 98.4\% respectively. For comparison, \citet{qu2021scone} reports an accuracy of 98.2\% for a 6-class classification task; our model demonstrates competitive performance despite the increased number of classes.

It should be noted, however, that a large number of samples (6048) are synthetically generated for classes that originally contained very few (\textless100) examples. This extensive oversampling may result in overlaps between synthetic and original samples in the feature space, which could make the results for those minority classes less reliable.

\begin{table*}
\centering
\caption{Performance metrics for multiclass classification, i.e., evaluation of how well the model distinguishes among multiple transient classes simultaneously.}
\label{tab:cat_class}
\begin{tabular}{ccccc} 
Metric & Training(\%) & Validation(\%) &Test(\%)  \\
\hline

Accuracy(\%) &  99.3 $\pm$ 0.2 &  98.4$\pm$ 0.1  & 98.3 $\pm$ 0.1 \\
Precision(\%) & 99.3 $\pm$ 0.2 & 98.3 $\pm$ 0.1 & 98.3 $\pm$ 0.0\\ 
Recall(\%) & 99.3 $\pm$ 0.2 & 98.4 $\pm$ 0.1 & 98.4 $\pm$ 0.1  \\
F1-score(\%) &  99.3 $\pm$ 0.2 & 98.3 $\pm$ 0.1 & 98.3 $\pm$ 0.1\\
\hline
\end{tabular}
\end{table*}

\subsection{Performance metrics for classifying other types of supernovae.}

In this section, we evaluate the model’s performance on binary classification tasks, distinguishing SNe~II and SNe~Ibc from all other types. This test is designed to assess whether the FPCA-based approach has potential applications in classifying Non-Ia transients. The corresponding performance metrics, along with the percentage representation of each class, are reported in Table~\ref{tab:met_oth_dis}.

The results show that the performance metrics for SNe~II are comparable to those for SNe~Ia, with accuracy, precision, and F1-score lower by only 2.9\%, 3.2\%, and 5.1\%, respectively. Since the differences are within 5\%, we conclude that this method is nearly equally effective for classifying both Type~I and Type~II SNe.  

For SNe~Ibc, however, the performance is notably weaker: accuracy, precision, and F1-score are lower by 1.8\%, 16.7\%, and 33.2\%, respectively. The dataset is highly imbalanced between SNe~Ibc and non-Ibc classes, which likely contributes to the poorer results. Oversampling techniques may improve performance but also increase the risk of overfitting; exploring such methods lies beyond the scope of this study.

It is also important to note that the penalty terms used in this analysis were derived from SNe~Ia light curves and applied uniformly across all transients. Using penalty terms specifically calibrated on clean samples of SNe~II or SNe~Ibc could further improve classification performance.

\begin{table*}
\centering
\caption{Performance metrics for classifying Non-Ia transients.}
\label{tab:met_oth_dis}
\begin{tabular}{lcccc} 
Metric & SN Ia & SN II & SN Ibc  \\
\hline
\% of  Objects& 54.2 & 35.1 &  6.0\\
Accuracy(\%) &  98.5 $\pm$ 0.0&  95.6 $\pm$ 0.1 & 96.7 $\pm$ 0.1   \\
Precision(\%) &   97.8 $\pm$ 0.1 & 94.6 $\pm$ 0.1 & 81.1 $\pm$ 1.8  \\ 
Recall(\%) & 99.5 $\pm$ 0.1 &  92.5 $\pm$ 0.2 & 54.8 $\pm$ 1.5   \\
F1-score(\%) &  98.6 $\pm$ 0.0 &  93.5 $\pm$ 0.1 & 65.4 $\pm$ 1.4 \\

\hline
\end{tabular}
\end{table*}

\subsection{Statistical Probabilistic Method}

In this section, we employ a statistical probability–based method to classify the transients. Such an approach is useful because, although ensemble methods (e.g., CatBoost) are generally more interpretable than black-box neural networks, they still introduce a degree of complexity and opacity due to the combined decision paths of multiple models, which can obscure intuitive understanding. The method used here is conceptually similar to the Nearest Neighbors algorithm, except that all training samples are included in determining the class of a given test sample. This method also serves as an independent validation check when classifying transients from observed data in real surveys.

This analysis is performed across three binary classification schemes: SNe~Ia vs.\ Non-Ia, SNe~II vs.\ Non-II, and SNe~Ibc vs.\ Non-Ibc. Since the number of other transient types in the dataset is fewer than 500 (out of more than 11,000), we do not attempt separate classifications for those classes.

The log-likelihood $\ln L_j({x}_{\text{test}})$ of an observed feature vector ${x}_{\text{test}}$  (with uncertainty $\sigma_{x_{\text{test}}} $) given the $j$-th training sample ${x}_{\text{train},j}$ (with uncertainty $\sigma_{x_{\text{train}},j} $)  is calculated as:

{\scriptsize
\[
\ln L_j({x}_{\text{test}}) = -0.5 \Bigg( 
\sum_{k=1}^{15} \frac{(x_{\text{train},j,k} - x_{\text{test},k})^2}{\sigma_{x_{\text{train}},j,k}^2 + \sigma_{x_{\text{test}},k}^2} 
+ \sum_{k=1}^{15} \ln(\sigma_{x_{\text{train}},j,k}^2 + \sigma_{x_{\text{test}},k}^2) 
\Bigg)
\]
}
This likelihood is computed for each  sample $j$ in the respective training set.

The total (unnormalized) scores for the SN Ia and Non-SN Ia classes, denoted $S_{\text{SN Ia}}$ and $S_{\text{Non-SN Ia}}$, are obtained by summing the exponentials of log-likelihoods for samples belonging to each class that satisfy a numerical stability threshold ($\ln L_j$\textgreater -700):
$$
S_{\text{SN Ia}} = \sum_{j \in \text{SNIa}} \exp\left(\ln L_j({x}_{\text{test}})\right)
$$
$$
S_{\text{Non-SN Ia}} = \sum_{j \in \text{Non-SN Ia}} \exp\left(\ln L_j({x}_{\text{test}})\right)
$$

Finally, the probability of an object being classified as a Type Ia supernova, $p(\mathrm{SN~Ia})$, is determined by normalizing these scores and incorporating a weighting factor $w$. In particular, the normalized scores are computed by dividing the total score by the number of training samples in each class:

$$
p(\text{SN Ia}) = \frac{\tilde{S}_{\text{SN Ia}}}{\tilde{S}_{\text{SN Ia}} + w \cdot \tilde{S}_{\text{Non-SN Ia}}}
$$

The weighting parameter $w$ is tuned by minimizing the total loss function
\[
\mathcal{L}_\text{total} = \sum_{\text{SNIa}} (1 - p) + \sum_{\text{Non-SNIa}} p,
\]
where $p = p(\text{SN Ia})$ is defined above.  

In a similar manner, these equations can be defined for SNe~II vs.\ Non-SNe~II and SNe~Ibc vs.\ Non-SNe~Ibc.

\begin{figure*}[h]
    \centering
    
        \includegraphics[width=\linewidth]{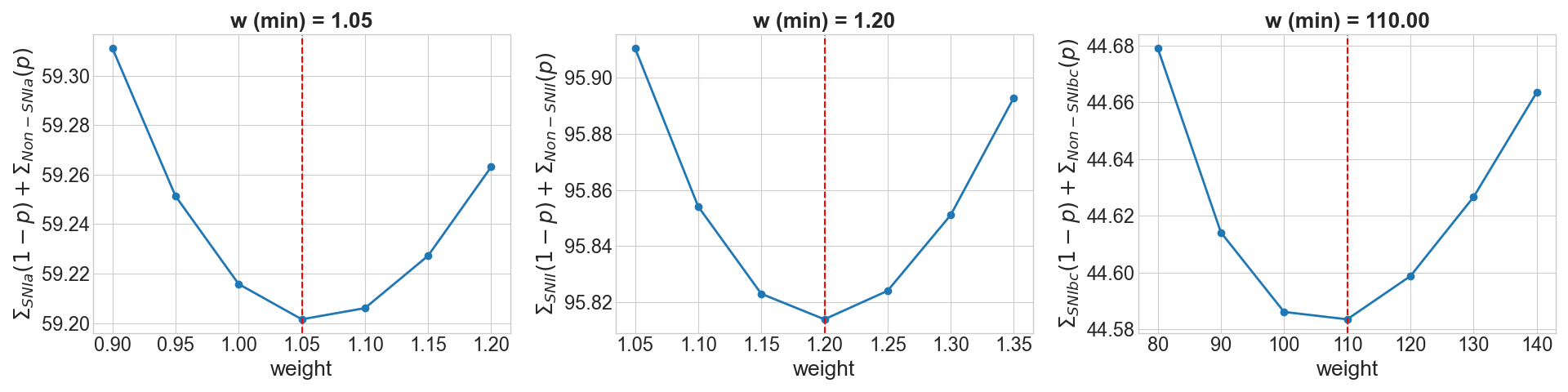}
    
    \caption{Validation loss for the function  $\Sigma_{\text{Type}}(1-p) + \Sigma_{\text{Non-Type}(p)}$, where Type refers to Ia (left), II (middle), and Ibc (right). The weighing parameter, w, is tuned such that the validation loss in minimized. The optimum value of w is used to report the accuracy on the test set. }
    \label{fig:loss_curve}
\end{figure*}

The validation loss curve is plotted in Figure~\ref{fig:loss_curve}. The optimal value of $w$, determined by tuning on the validation set, is shown as the title of each subplot. Using this value of $w$ and a threshold probability of 0.5, the overall accuracy of the model is evaluated on the test set to enable a meaningful comparison with the machine learning algorithm.

\begin{figure*}
  \centering
 
    \includegraphics[width=1.0\linewidth]{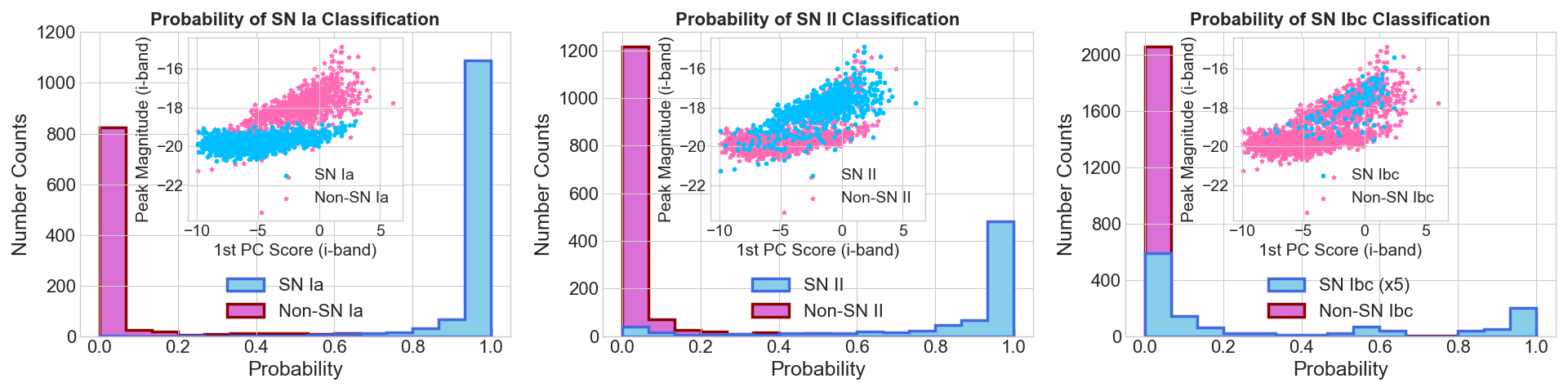}
 
  \caption{Probability distributions of test samples computed using the statistical method. The left, middle, and right panels correspond to SNe~Ia, SNe~II, and SNe~Ibc, respectively. The target and non-target classes are shown in blue and purple, respectively. The inset scatter plot displays the distributions of peak absolute magnitude versus the first principal component (PC) score in the $i$-band for the corresponding binary classification problem.}
  \label{fig:prob}
  
\end{figure*}

 We plot the histograms for each of the three cases in Figure~\ref{fig:prob}. The samples from the target class are shown as blue histograms, while the corresponding samples from the non-target class are shown in purple. These histograms represent the probability distributions of test samples classified as the target class. The inset scatter plot displays the distributions of two parameters (peak absolute magnitude vs.\ first PC score) in the $i$-band.  

The overall accuracy obtained using the two methods is 98.5\% (Ia), 94.6\% (II), and 96.7\% (Ibc) for the machine learning approach, versus 95.7\% (Ia), 91.3\% (II), and 95.7\% (Ibc) for the statistical method, with a maximum difference of 3\%. This demonstrates a partial separation of SNe~Ia (and other transients) in the parameter space encoded by the feature vector set considered in this study.

\section{Applications to DECam data}
\label{applications}

\subsection{The Dark Energy Survey Supernova Program}

The spectroscopic sample of light curves from the first five years of the Dark Energy Survey Supernova Program \citep{sanchez2024dark}  provides a valuable benchmark for evaluating the domain adaptability of our method, i.e., its ability to maintain performance when applied to data with properties differing from the training set. Domain adaptability is crucial because surveys often have varying or partially unknown observational characteristics, and the classifier must generalize robustly across varying observational conditions.

Our selection criteria include observations with reasonable signal-to-noise ratios (magnitude error \textless 0.5) and at least five observations in at least one filter, resulting in a sample of 456 spectroscopically confirmed transients and variables. The DES data is available in four filters (g, r, i, z); we retrain the model using PLAsTiCC simulations, discarding the y-band parameters. We then apply our model to these 456 spectroscopically confirmed transients to evaluate its classification performance and assess its domain adaptability.

 \subsubsection{Ensemble-based boosting}
The model trained with the reduced feature vector set, using the previously tuned hyperparameter values from the PLAsTiCC validation dataset, achieves an accuracy of 89.7\% and a precision of 95.9\%. Out of 338 spectroscopically confirmed SNe~Ia in the DES dataset, 306 were correctly identified, and out of 118 Non-SNe~Ia, 105 were correctly classified. No systematic trend is observed in the misclassifications. Since the accuracy difference between the training dataset (PLAsTiCC) and the DES data is less than 10\%, we can conclude that the model's performance is reasonably satisfactory on DES data.

The redshift distribution of the PLAsTiCC dataset differs significantly from that of the DES spectroscopic data, with the latter dominated by low-redshift objects. We investigated the effect of matching the redshift distribution between training and test samples on the model's performance. This was done by dividing the test samples into redshift bins, determining the fraction of samples in each bin, and selecting the same fraction of samples from the training dataset within the same redshift ranges. However, this reduces the size of the training data (unless artificial samples are generated), because the fraction of samples within specific redshift ranges can vary widely between datasets. Using this approach, we selected 2390 out of the original 11158 training samples (\(\approx 20\%\)). The resulting training set contains only 28.3\% SNe~Ia, whereas the DES dataset has 74.1\% SNe~Ia. The overall accuracy and precision obtained using this matched redshift distribution are 90.1\% and 96.5\%, respectively.  

Since the performance metrics of both methods are statistically indistinguishable, we conclude that an ensemble-based classifier trained using FPCA scores is largely insensitive to the redshift distribution of the training data. This is a major advantage of the method, as the model does not need to be retrained on new simulations before being applied to observed data from a particular survey.

\subsubsection{Statistical Probabilistic Method}

We apply the method outlined in Section~5.13 to classify the DES samples, using the matched redshift distribution for selecting the training sample as described in the previous section, since matching the redshift distribution provides slightly better performance than using an unmatched distribution.
 The weighting parameter (the only hyperparameter for this method) is retained from the PLAsTiCC validation set, and probabilities are calculated based on the separation in parameter space for the $g$, $r$, $i$, $z$ band features. Out of 338 spectroscopically confirmed SNe~Ia in the dataset, 308 were correctly identified, and out of 118 Non-SNe~Ia, 101 were correctly classified. The overall accuracy and precision obtained using this method are 89.7\% and 94.8\%, respectively. The performance of both methods is nearly identical, with CatBoost slightly outperforming the statistical method by roughly 1\%.  

Although the two methods are based on different underlying principles, their comparable performance on this dataset suggests that both are robust. Consequently, objects consistently classified as Type~Ia supernovae by both methods are likely to form a more reliable subsample with reduced contamination, suitable for high-precision cosmological analyses.

The results for all cases of DES transients are tabulated in Table~\ref{tab:spec}. This application demonstrates the high domain adaptability of FPCA summary statsitics, in contrast to previous state-of-the-art approaches such as ParSNIP, which require retraining when applied to a new dataset.

\begin{table}
\centering
\caption{Performance metrics for the model applied to DES spectroscopic sample.}
\label{tab:spec}
\begin{tabular}{ccc} 
Method & Accuracy(\%) & Precision(\%)   \\
\hline

ML(unmatched z) &  89.7  &  95.9 \\
ML(matched z) & 90.1  & 96.5 \\ 
Statistical (unmatched z) & 88.6  & 94.8   \\

\hline
\end{tabular}
\end{table}

\subsection{DECam Deep Drilling Field}

We apply this model to the light curves from the DECam deep drilling field \citep{graham2023deep} to identify potential SNe~Ia among the detected transients. A key challenge is to remove variable stars, which are not represented in the PLAsTiCC dataset. As a primary filter, we discard samples with a standard deviation in magnitude less than 0.2. As a secondary check, we use the goodness-of-fit, discarding light curves with $\frac{\chi^2}{\nu} < 5$, resulting in a sample of 156 transients.  

Both CatBoost and the statistical probability method are applied to these 156 transients, and the predictions are compared with the potential SNe~Ia candidates listed in the follow-up characterization paper \citep{graham2024deep}. \citet{nugent2002k} proposed a parametric relation between amplitude and light curve span, which \citet{graham2024deep} used to derive a starting set of 100 potential SN~Ia candidates. These 100 light curves were then fitted using \texttt{sncosmo} \citep{barbary2016sncosmo} and the SALT2 model \citep{guy2005salt}, assuming all were SNe~Ia. Out of these, 62 candidates yielded reasonable SALT2 parameter fits. We compare our findings with these 62 potential SNe~Ia candidates.

\subsubsection{Ensemble-based Boosting}

Out of the 156 transients, 97 are classified as Type~Ia SNe by the CatBoost model trained as described above. We then conducted a further inspection of the light curves to identify and exclude problematic cases, such as those lacking data before or after peak brightness or having observations in only one filter. This refinement, performed in this work, resulted in a sample of 56 Type~Ia supernova candidates. After applying the cuts based on standard deviation and goodness-of-fit, as described previously, along with the requirement of at least five observations in at least one filter, 47 out of the 62 potential SNe~Ia candidates identified by SALT2 fitting in the follow-up study \citep{graham2023deep} remain. Among these, 35 were identified as Type~Ia SNe by the CatBoost model using a threshold probability of 0.5.

\subsubsection{Statistical Probabilistic Method}

Applying the statistical probabilistic model to the same set of 156 transients, 59 were classified with a probability greater than 0.5 of being SNe~Ia. Among these 59 candidates, 30 are listed among the 62 potential SNe~Ia candidates in \citet{graham2023deep}, and 38 are classified as SNe~Ia by the ensemble model. Since the predictions for this intersected sample of 38 candidates are identical for both models, we conclude that these constitute a high-confidence sample of SNe~Ia, with the full list given in Table~\ref{tab:decam_ddf}.

\begin{table*}
\centering
\caption{List of high-confidence SNe~Ia candidates among the DECam DDF transients, defined as those classified as Type~Ia by both the CatBoost and statistical probabilistic methods.}
\label{tab:decam_ddf}
\begin{tabular}{|c|c|c|c|c|c|c|c|} 
\hline

DC21bohl & DC21bpls & DC21bqmr & DC21bxvcb & DC21cgfrp & DC21cozcn   & DC21cwtlp  & DC21cyddn   \\
\hline
DC21cyks &DC21dpwcq & DC21dqpcm  &DC21dqqdr &DC21dsocp  & DC21dtwb &DC21esjb  & DC21fmlu\\
\hline
 DC21fsdd & DC21ghpw  & DC21hpdc & DC21hunj & DC21ktp & DC22dulzc  & DC22gtowb & DC22guhrf \\
\hline
DC22hvtwp & DC22hvuzr & DC22hxihw  & 
DC22ikjao   &DC22jcyuz  & DC23jjloo & DC23jmqzz & DC23jqitt \\
\hline

DC23jwhoq & DC23jwnzq  & DC23kpigs  &  DC23kvgpp  &  DC23kwpaw & DC23kzwjr &  &  \\
\hline

\end{tabular}
\end{table*}

\subsection{DECam Survey of Intermediate Redshift Transients (DESIRT)}

The DECam Survey of Intermediate Redshift Transients \citep{flaugher2012status} is designed to detect transients at redshifts between 0.2 and 0.8, bridging the gap between local (low-$z$) and distant (high-$z$) transient surveys. This imaging and photometric survey primarily targets both Type~Ia and Core Collapse SNe to improve cosmological distance measurements, study star formation history, and investigate progenitor environments and host galaxies of the exploding transients.

We applied our model trained on the PLAsTiCC dataset to 69 transient candidates detected in 2021 that were not within 0.3~arcsec of a previously identified galaxy and had at least five data points in three filters ($g$, $r$, and $z$). Since PLAsTiCC contains photometry in five bands while DESIRT provides only three, we retrained the model using PLAsTiCC data with the $i$ and $y$-band features removed, and report the results in Table~\ref{tab:met_g_i}. Comparing the second and third columns, removing the two bands results in a decrease of approximately 1\% in all performance metrics, demonstrating that the model can be reliably applied to survey data with only $g$, $r$, and $z$ photometry.

\begin{figure*}
    \centering
    \subfloat{\includegraphics[width=0.45\textwidth]{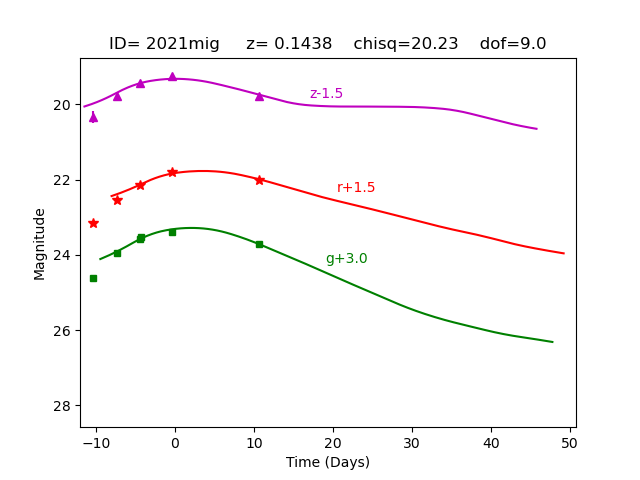}}
    \hfill
    \subfloat{\includegraphics[width=0.45\textwidth]{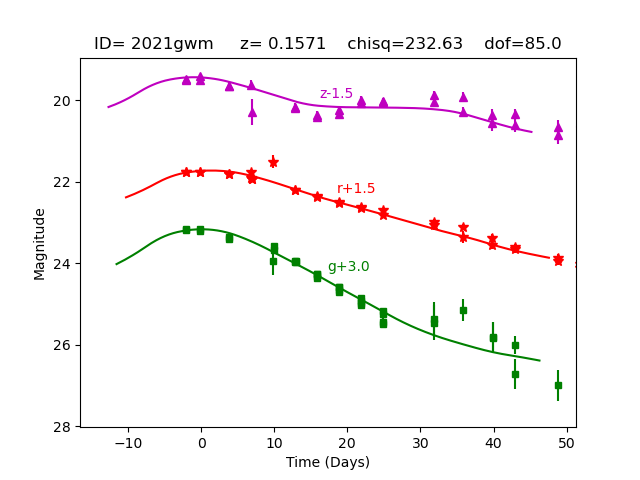}}

    \vskip 0.5cm 

    \subfloat{\includegraphics[width=0.45\textwidth]{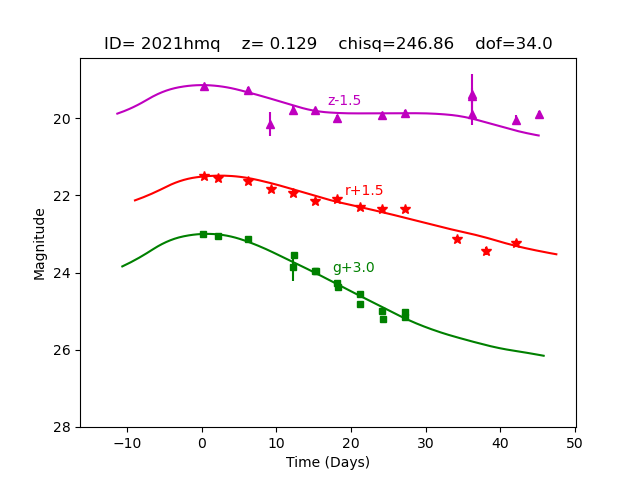}}
    \hfill
    \subfloat{\includegraphics[width=0.45\textwidth]{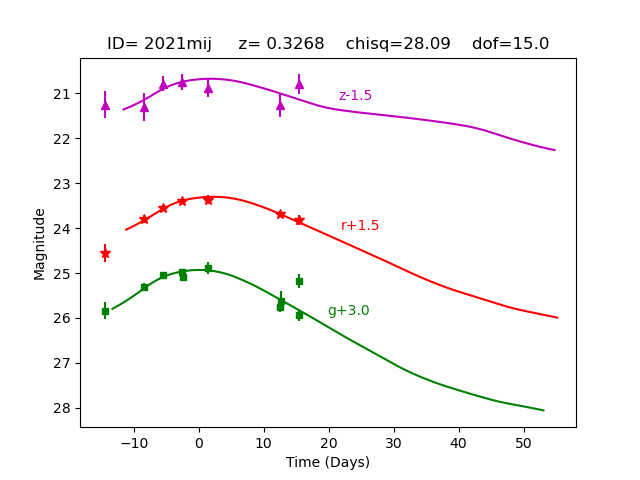}}

    \caption{Examples of four light curves from DESIRT transients (supernova candidates) classified as Type~Ia supernovae by both the CatBoost and statistical probabilistic methods.}
    \label{fig:des_tra_snIa}
\end{figure*}

Out of the 69 transients, 37  were classified as Type~Ia by CatBoost, and a visual inspection of the light curves was consistent with these classifications. The statistical probability-based method yielded similar results, identifying 38 transients as SNe~Ia. The intersection of the two methods includes 34 transients, representing a robust subsample of high-confidence Type~Ia candidates. Figure~\ref{fig:des_tra_snIa} presents four representative examples from this intersected set.

It is worth noting that this is a preliminary analysis, and a detailed paper on DESIRT transient detection and classification (Hu et al., in prep.) is in preparation.

\begin{table*}
\centering
\caption{Performance metrics of the model on the PLAsTiCC dataset when the $i$ and $y$ bands are removed from the feature set, demonstrating its reliability when applied to survey data with photometry in only the $g$, $r$, and $z$ bands.}
\label{tab:met_g_i}
\begin{tabular}{ccccc} 
Metric & 5-bands & 3-bands  \\
\hline

Accuracy(\%) &   98.5 $\pm$ 0.0 & 97.4 $\pm$ 0.1 \\
Precision(\%) &    97.8 $\pm$ 0.1 & 97.0 $\pm$ 0.1 \\ 
Recall(\%) & 99.5 $\pm$ 0.1 & 98.3 $\pm$ 0.1  \\
F1-score(\%) & 98.6 $\pm$ 0.0 & 97.7 $\pm$ 0.1\\

\hline
\end{tabular}
\end{table*}

\section{conclusions and discussions} \label{sec:conclusions}

In this paper, we have demonstrated the feasibility of photometric identification of Type~Ia SNe using functional principal component analysis (FPCA) light-curve modeling combined with ensemble learning (CatBoost). This approach is novel in leveraging FPCA scores—previously shown to correlate with physical properties of SNe~Ia—to distinguish them from other transients. The FPCA basis functions are derived in a data-driven manner, providing a flexible and empirical modeling framework capable of capturing higher-order modes of variability in the data, in contrast to methods based on rigid theoretical assumptions such as SALT2. Compared to state-of-the-art methods like Gaussian process regression or convolutional neural networks, our approach is computationally simpler and does not require GPU-intensive resources.

Our analysis incorporates all transients detected within 10 days past maximum light with at least five observations in one filter, regardless of type or redshift. Light curves are fitted by minimizing the residuals with respect to the observations, including penalty terms derived from clean samples, to mitigate the effect of outliers. Interband peak-time offsets are incorporated to model bands with limited observations, provided at least one band is well-sampled. Feature importance analysis shows that the peak absolute magnitude and first PC scores in the $r$, $i$, and $z$ bands are the most discriminative for SNe~Ia. Overall, the model achieves accuracy, precision, and AUC of 98.5\%, 97.8\%, and 0.997, respectively,  comparable to state-of-the-art-methods. Performance is slightly reduced for high-redshift transients ($z \sim 1$) or light curves with narrow time coverage ($\sim$20 days). The performance remains robust for typical photometric uncertainties expected in surveys such as LSST ($\sigma$=0.1). The FPCA scores also enable robust classification of other transient types, such as SNe~II and SNe~Ibc.

We developed a complementary statistical probabilistic model based on an exponential metric in parameter space based on the same FPCA scores. This method provides comparable results (95.7\%) to the ensemble approach (98.5\%), serving as an independent validation.

Application to DECam-based surveys—including DES YR5, DECam DDF, and DESIRT—demonstrates the method’s high domain adaptability. Using the DES YR5 spectroscopic sample as a benchmark, we achieve $\sim$90\% accuracy despite significant differences in redshift distributions compared to the training set. Unlike other state-of-the-art methods (e.g., ParSNIP, Avocado) that require retraining on new datasets, our approach performs reliably without retraining and is insensitive to the redshift distribution of the training sample. Predictions from both methods show strong agreement for DECam DDF and DESIRT transients, and the intersection of the two independent methods provides a high-confidence SNe~Ia sample suitable for cosmological applications.

Overall, our data-driven light curve modeling combined with automated machine learning demonstrates robust performance across multiple surveys and transient types. This makes it a promising tool for analyzing millions of light curves from upcoming large-scale surveys such as LSST and Roman. The FPCA scores can also serve as informative features in future cosmological analyses, enabling deeper insights into the fundamental nature of the Universe.

\section*{Acknowledgements}
We thank Peter Brown for providing valuable insights on the UV templates of the SALT model for SNe~Ia, as well as guidance regarding the PLAsTiCC dataset. We also thank Nick Suntzeff for his helpful comments on the manuscript.

\section*{Data and Code Availability}
The analysis pipeline and supporting code developed for this work will be made publicly available on GitHub upon acceptance of the manuscript.

\bibliography{paper}{}
\bibliographystyle{aasjournal}

\appendix

\section{Different ML Models}

\begin{figure*}
    \centering
        \includegraphics[width=0.9\linewidth]{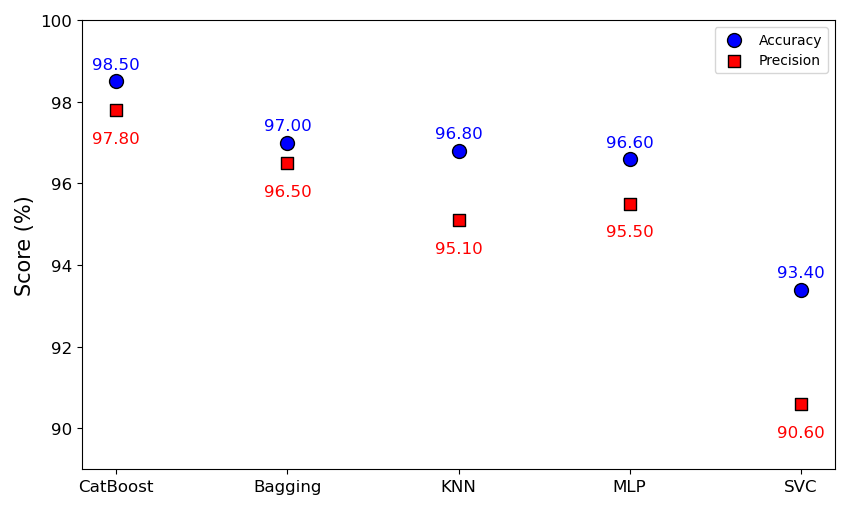}       
    \caption{Accuracy and precision obtained using five different machine learning models. The highest accuracy and precision are achieved with the CatBoost classifier.}
    \label{fig:diff_ml}
\end{figure*}

In this section, we compare the performance of different machine learning models for distinguishing SNe~Ia from other types of transients. Figure~\ref{fig:diff_ml} shows the accuracy and precision for four alternative models alongside CatBoost (used in this study), with accuracy and precision indicated by blue circles and red squares, respectively. Among these five models, only CatBoost can handle missing values directly; for the other four models, missing values were imputed using a nearest neighbors algorithm before being input to the classifiers.

Briefly, the models are as follows: Bagging is similar to boosting but connects constituent decision trees in parallel rather than sequentially; k-Nearest Neighbors (kNN) classifies samples based on the closest neighbors in parameter space; Multilayer Perceptron (MLP) is a feedforward neural network with neurons whose weights are adjusted to minimize the loss; and Support Vector Classification (SVC) finds an optimal hyperplane separating the classes. As shown in the figure, both accuracy and precision exceed 95\% for all methods except SVC, with CatBoost outperforming the others.

\section{Linear Dependence of redshift}

\begin{figure*}
    \centering
    \includegraphics[width=\textwidth]{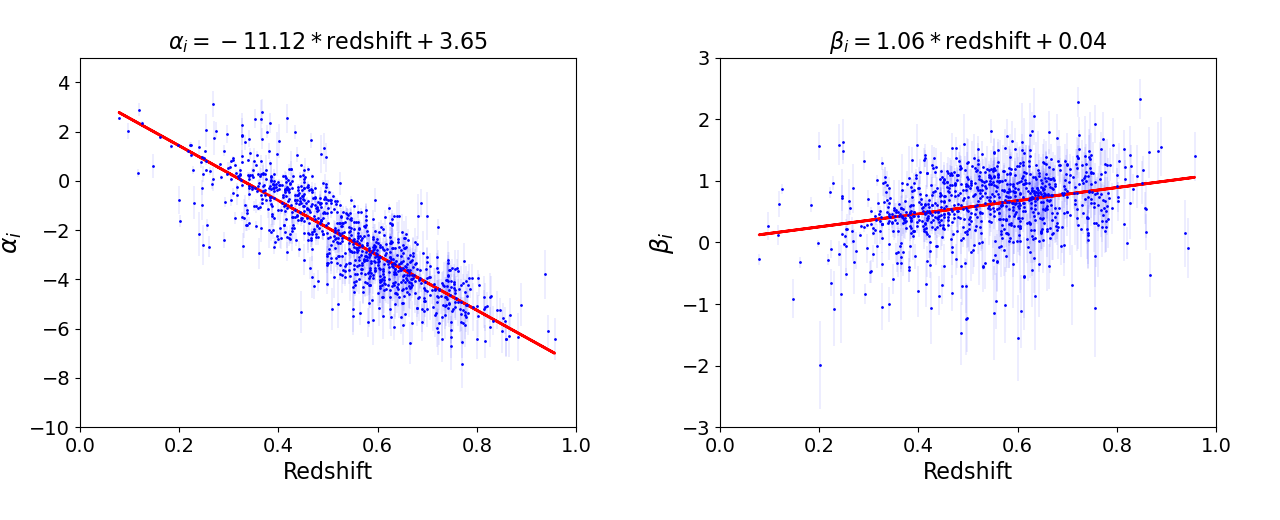}
    \caption{. Distribution of the first FPCA coefficients (left panel) and the second FPCA  coefficients (right panel) with redshift. Blue dots denote the data points, while the red line shows the best-fit linear relation obtained from least squares minimization.}
    \label{fig:data_red_dep}
\end{figure*}

To justify the choice of modeling the FPCA penalty coefficients as linear functions of redshift, we examined their empirical behavior across the training sample. As shown in Figure~\ref{fig:data_red_dep}, both sets of coefficients ($b_{1i}, b_{2i}$ for the first penalty score and $c_{1i}, c_{2i}$ for the second) exhibit an approximately linear dependence on redshift. These coefficients are modeled as: $\alpha_i = b_{1i} z + b_{2i}$ and $\beta_i = c_{1i} z + c_{2i}$, where $i$ indexes the photometric band.

Similar linear trends are observed across all photometric bands considered in this work, providing empirical support for the parameterization adopted in Equation~\ref{eqn:prior_scores}.

\section{Data Visualization}

In this section, we explore two methods of data compression suitable for visualization. For very high-dimensional feature spaces, the compressed data vectors can also be used for model training to reduce computational complexity. In our study, since model training and testing require less than a minute, data compression does not provide a computational advantage. Nevertheless, we train the model using two compressed feature vectors and compare the performance with the base classifier.

\subsection{Principal Component Analysis}

Principal Component Analysis (PCA) \citep{hasan2021review, salem2019data} is a linear dimensionality reduction technique commonly used in statistics and machine learning to compress data while preserving variance. The principal components (PCs) are constructed to be orthogonal, with the first PC capturing the largest variance, followed sequentially by higher-order components. We construct a 2-component PCA from our original 15-feature vector and show the distribution in the left panel of Figure \ref{fig:data_vis}.. For a quantitative evaluation, we train and test the model using these two features, with results shown in the second column of Table \ref{tab:data_vis}. For comparison, the performance using all 15 features is shown in the first column. All metrics exceed 90\%, except precision (89.3\%).  The accuracy and precision decrease by about 6\% and 9\% respectively compared to the full feature set.

\subsection{UMAP}

\begin{figure*}
    \centering
    \includegraphics[width=\textwidth]{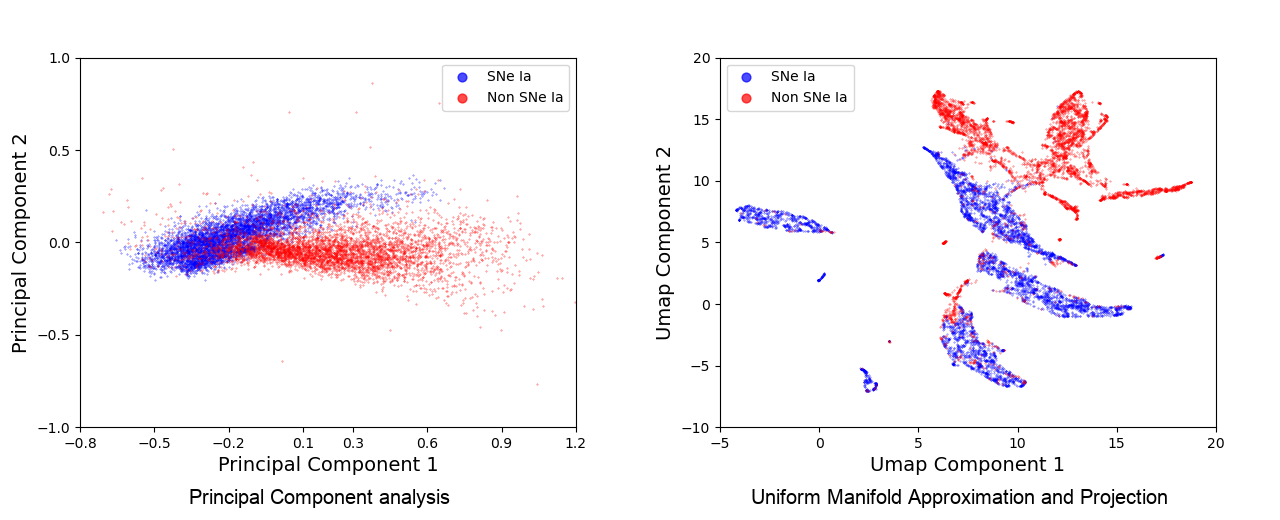}
    \caption{Distribution of the first two parameters, which capture the highest variance, visualized using two different data compression methods.}
    \label{fig:data_vis}
\end{figure*}

UMAP \citep{mcinnes2018umap} (Uniform Manifold Approximation and Projection) is a non-linear dimensionality reduction technique. It constructs a graph of high-dimensional data points based on their nearest neighbors, and then calculates the connectivity of each point to others, forming a ``fuzzy topological structure" of the dataset. While PCA captures global variance, UMAP preserves local structures, which may lead to overfitting. The 2-component UMAP embedding is shown in the right panel of Figure \ref{fig:data_vis}, and the corresponding performance metrics are presented in the fourth column of Table \ref{tab:data_vis}. The performance is comparable to the default method using all 15 features and notably better than the PCA-based results.

\begin{table*}
\centering
\caption{Performance metrics obtained using 2-component feature vectors for two different data visualization methods.}
\label{tab:data_vis}
\begin{tabular}{lcccc} 
Metric & Base &PCA  & U-MAP   \\
\hline
Explained Variance & 100.0 & 81.0 & -\\
Accuracy(\%) & 98.5 $\pm$ 0.0 & 92.0 $\pm$ 0.2 & 95.1 $\pm$ 0.2   \\
Precision(\%) & 97.8 $\pm$ 0.1 & 89.3 $\pm$ 0.2 &  92.9$\pm$ 0.3  \\ 
Recall(\%) & 99.5 $\pm$ 0.1 &  97.3 $\pm$ 0.2 &  98.7$\pm$ 0.3   \\
F1-score(\%) &  98.6 $\pm$ 0.0 & 93.1 $\pm$ 0.2 & 95.7 $\pm$ 0.2 \\

\hline
\end{tabular}

\end{table*}

\end{document}